\documentclass[aps,prb,twocolumn,showpacs,preprintnumbers,amsmath,amssymb,groupedaddress,noshowkeys]{revtex4}
\usepackage{txfonts}
\usepackage[dvipdfm]{graphicx}
\usepackage{bm}
\def\rnum#1{\expandafter{\romannumeral #1}} 
\def\Rnum#1{\uppercase\expandafter{\romannumeral #1}}

\newfont{\bg}{cmr10 scaled\magstep4}

\newcommand{\bigzerou}{\smash{\lower1.8ex\hbox{\bg 0}}}

\begin{document}

\title{Anomalous Josephson effect induced by spin-orbit interaction
and Zeeman effect in semiconductor nanowires}
\author{Tomohiro Yokoyama}
\email[E-mail me at: ]{tomohiro.yokoyama@riken.jp}
\affiliation{Center for Emergent Matter Science, RIKEN institute,
2-1 Hirosawa, Wako, Saitama 351-0198, Japan}

\author{Mikio Eto}
\affiliation{Faculty of Science and Technology, Keio University,
3-14-1 Hiyoshi, Kohoku-ku, Yokohama 223-8522, Japan}

\author{Yuli V.\ Nazarov}
\affiliation{Kavli Institute of Nanoscience, Delft University of Technology,
Lorentzweg 1, 2628 CJ, Delft, The Netherlands}
\date{\today}

\begin{abstract}
We investigate theoretically the Josephson junction of
semiconductor nanowire with strong spin-orbit
(SO) interaction in the presence of magnetic field.
By using a tight-binding model, the energy levels
$E_n$ of Andreev bound states are numerically calculated
as a function of phase difference $\varphi$ between
two superconductors in the case of short junctions.
The DC Josephson current is evaluated from
the Andreev levels.
In the absence of SO interaction, a $0$-$\pi$ transition
due to the magnetic field is clearly observed.
In the presence of SO interaction, the coexistence of
SO interaction and Zeeman effect results in
$E_n (-\varphi ) \ne E_n (\varphi )$, where
the anomalous Josephson current flows even at $\varphi =0$.
In addition, the direction-dependence of critical current
is observed, in accordance with experimental results.
\end{abstract}
\maketitle

\section{INTRODUCTION}
\label{sec:Intro}

The spin-orbit (SO) interaction in narrow-gap semiconductors,
e.g., InAs and InSb,~\cite{Winkler} attracts a lot of interest in
recent studies.
The SO interaction gives a possibility of electrical spin manipulation,
which is a great advantage for spintronic devices.~\cite{Zutic}
For conduction electrons in direct-gap semiconductors,
the SO interaction is expressed as
\begin{equation}
H_{\rm SO} =\frac{\lambda}{\hbar} \bm{\sigma} \cdot
\left[\bm{p} \times \bm{\nabla} V(\bm{r}) \right],
\label{eq:SO}
\end{equation}
where $V(\bm{r})$ is an external potential and $\bm{\sigma}$
indicates the electron spin $\bm{s}=\bm{\sigma}/2$.
In experiments of quantum well using such materials,
strong SO interaction was reported.~\cite{Nitta,Grundler,Yamada}
For an external electric field ${\cal E}$ perpendicular to
the quantum well, the substitution of $V(\bm{r})=e {\cal E} z$
into Eq.\ (\ref{eq:SO}) yields
\begin{equation}
H_{\rm SO} =\frac{\alpha}{\hbar} (p_y\sigma_x-p_x\sigma_y),
\label{eq:Rashba}
\end{equation}
which is called the Rashba interaction. Here, the coupling
constant $\alpha=e {\cal E} \lambda$ is tunable by
an electric field, or a gate voltage.

The development of fabrication technique enables to
construct various quantum systems with SO interaction.
Particularly, semiconductor nanowires of InAs and InSb
are investigated intensively,
in which quantum point contacts and quantum dots can be
formed.~\cite{Fasth,Pfund,Nadj-Perge,Nadj-Perge2,Nadj-Perge3,Schroer}
Indeed, the electrical manipulation of single electron spins
was reported for quantum dots fabricated on the
nanowires.~\cite{Nadj-Perge2,Nadj-Perge3,Schroer}
In recent studies, the nanowire-superconductor hybrid
systems were studied for searching the Majorana
fermions.~\cite{Mourik,Rokhinson,Das,Deng}
The DC Josephson effect was also studied when
the nanowires are connected to two superconductors
(S/NW/S junctions).~\cite{Doh,Dam,Nilsson}

The Josephson effect is one of the most fundamental
phenomena concerning quantum phase.
In a Josephson junction, the supercurrent flows when
the phase difference $\varphi$ between two superconductors is
present. For the junction using normal metals or semiconductors,
the electron and hole in the normal region are coherently
coupled to each other by the Andreev reflections at
normal / superconductor interfaces.~\cite{Andreev}
The Andreev bound states, which have discrete
energy levels $E_n$ (Andreev levels), are formed in the normal
region around the Fermi level within the superconducting
energy gap $\Delta_0$.~\cite{Beenakker1,NazarovBlanter}
The Cooper pair transports via the Andreev bound states.
For short junctions, where a distance $L$ between
two superconductors is much smaller than
the coherent length $\xi$ in the normal region,
the Josephson current $I (\varphi )$ is determined by
the Andreev levels.~\cite{Beenakker1,Bardeen,Furusaki1,Furusaki2}
$\xi=\hbar v_{\rm F} / (\pi \Delta_0) \equiv \xi_0$ for ballistic
systems and $\xi =(\xi_0 l_0)^{1/2}$ for diffusive ones, where
$v_{\rm F}$ is the Fermi velocity and $l_0$ is the mean free path.
For the transmission probability $T_n$ of conduction channel
$n$ ($=1,2,\cdots,N$) in the normal region, the current
is written as
\begin{equation}
I (\varphi )=\frac{e\Delta_0}{2\hbar} \sum_{n=1}^{N}
\frac{T_n \sin \varphi }{[1-T_n \sin^2 (\varphi /2)]^{1/2}}.
\label{eq:Josephson}
\end{equation}
Here, the current satisfies $I(-\varphi) = -I(\varphi)$.
In the limit of low transparent junction, the current
in Eq.\ (\ref{eq:Josephson}) becomes
$I(\varphi) \simeq I_0 \sin \varphi$ with
$I_0 \equiv e\Delta_0 /(2\hbar) \sum_n T_n$.

In superconductor / ferromagnet / superconductor (S/F/S)
junctions, the oscillation of critical current accompanying
a $0$-$\pi$ transition was observed as a function of
the thickness of ferromagnet.~\cite{Buzdin1,Kontos,Ryazanov,Oboznov}
The $0$- and $\pi$-states mean that the free energy is
minimal at $\varphi = 0$ and $\pi$, respectively.
The $0$-$\pi$ transition is caused by the interplay between
the spin-singlet correlation of the superconductivity and
the exchange interaction in ferromagnet.
The exchange interaction makes the spin-dependent
phase shift in the propagation through the ferromagnet.
Since the Andreev bound state consists of
a right-going (left-going) electron with spin $\sigma$
and a left-going (right-going) hole with spin $-\sigma$,
the phase shift modulates the Andreev levels.
When the length of ferromagnet is increased,
the $0$-$\pi$ transition takes place at the cusps of
critical current.~\cite{Oboznov}
A similar transition was observed recently in S/NW/S
junctions with fixed length when the Zeeman splitting was
tuned by applying a magnetic field.~\cite{private}

The effect of SO interaction in the Josephson junctions
is an interesting subject,
where many phenomena were predicted,
e.g., fractional Josephson effect~\cite{FuKane} and
anomalous supercurrent.~\cite{Buzdin2}
The fractional Josephson effect is a $4\pi$ periodicity of
current phase relation, $I(\varphi) \sim \sin (\varphi /2)$,
which is a property of Majorana fermions induced by
the SO interaction in the superconducting region.
The anomalous supercurrent is a finite supercurrent
at zero phase difference, $I (\varphi =0) \ne 0$,
which is induced by the breaking of symmetry of
current phase relation. The symmetry breaking
is attributed to the existence of SO interaction
and magnetic field in the normal region.

In the present study, we focus on the anomalous Josephson
current. The DC Josephson current with SO interaction in
the normal region was investigated theoretically by
a lot of groups, for normal metal with magnetic impurities,~\cite{Buzdin2}
two-dimensional electron gas (2DEG) in semiconductor
heterostructures,~\cite{Bezuglyi,Liu2,Liu1,Liu3,Malshukov1,Malshukov2,Malshukov3}
open quantum dots,~\cite{Beri}
quantum dots with tunnel
barriers,~\cite{DellAnna,Zazunov,Dolcini,Karrasch,Droste,Padurariu,Brunetti}
carbon nanotubes,~\cite{Lim}
quantum wires or nanowires,~\cite{Krive1,Krive2,Cheng,YEN}
quantum point contacts,~\cite{Reynoso1,Reynoso2}
topological insulators,~\cite{Tanaka}
and others.~\cite{Chtchelkatchev2}
The SO interaction breaks the spin-degeneracy of
Andreev levels when the time reversal symmetry is
broken by the phase difference $\varphi \ne 0$
even in the absence of magnetic field.~\cite{Beri,Chtchelkatchev2}
The splitting due to the SO interaction is obtained in
the long junctions, $L \gg \xi$
(or intermediated-length junctions, $L \gtrsim \xi$).
In the short junctions, however, the spin-degeneracy of
Andreev levels holds.~\cite{Beri,Chtchelkatchev2}
In both cases, the relation of $I(-\varphi) = -I(\varphi)$
is not broken, which means no supercurrent at $\varphi =0$.

In the presence of magnetic field, the SO interaction
modifies qualitatively the current phase relation.
Then, the anomalous Josephson current is
obtained.~\cite{Buzdin2,Liu1,Liu3,Zazunov,Brunetti,Krive1,Reynoso1,Reynoso2,YEN}
The anomalous current flows in so-called $\varphi_0$-state
in which the free energy has a minimum at
$\varphi =\varphi_0$ ($\ne 0,\pi$).~\cite{comphi0}
The anomalous Josephson current was predicted
when the length of normal region $L$ is longer than
or comparable to the coherent length $\xi$.
Krive {\it et al.} derived the anomalous current for
long junctions with a single conduction channel.~\cite{Krive1}
Reynoso {\it et al.} found the anomalous current through
a quantum point contact in the 2DEG for
$L \gtrsim \xi$.~\cite{Reynoso1,Reynoso2}
They discussed an influence of spin polarization
induced around the quantum point contact with
SO interaction~\cite{EtoQPC} on the Josephson current.
They also showed the direction-dependence of critical
current when a few conduction channels take part in the transport.
In the experiment on the S/NW/S junctions,~\cite{private}
the direction-dependent supercurrent was observed for
samples of $L \gtrsim \xi$ in a parallel magnetic field to
the nanowire, besides the above-mentioned $0$-$\pi$ transition.
This should be ascribable to the strong SO interaction
in the nanowires although the anomalous Josephson
current was not examined.


In our previous paper, we investigated theoretically
the DC Josephson effect in semiconductor
nanowires with strong SO interaction
in the case of short junction.~\cite{YEN}
We examined a simple model with single scatterer
to capture the physics of $0$-$\pi$ transition
and anomalous Josephson effect. In our model,
both elastic scatterings by the impurities and
SO interaction in the nanowire were represented by
the single scatterer.
The Zeeman effect by a magnetic field shifts
the wavenumber as
$k_\pm^>= k_{\rm F} +(E \pm E_{\rm Z})/(\hbar v_{\rm F})$
for $k>0$ and
$k_\pm^<=-k_{\rm F} -(E \pm E_{\rm Z})/(\hbar v_{\rm F})$
for $k<0$.~\cite{Chtchelkatchev1}
$E$ is an energy measured from the Fermi level.
$E_{\rm Z} \equiv |g \mu_B B|/2$ is the Zeeman energy.
The orbital magnetization is neglected in the nanowire.
The Fermi velocity $v_{\rm F}$ is independent of channels.
When $k>0$ ($k<0$), the electron and hole move to
the right (left) and left (right), respectively.
The propagation of electron with spin $\sigma =\pm$
and hole with $\sigma =\mp$ acquires the phase
$\pm \theta_{B} =\pm |g\mu_{\rm B}B|L/(\hbar v_{\rm F})$.
The term of $2EL /(\hbar v_{\rm F})$ is safely
disregarded for short junctions. The terms of
$k_{\rm F}$ are canceled out by each other.
The simple model showed the oscillation of
critical current with increase of $\theta_B$.
The $0$- and $\pi$-states are realized when
$\theta_B \sim 0$ and $\pi$, respectively.
Around $\theta_B =\pi /2$, the $0$-$\pi$ transition
takes place.
In the presence of SO interaction, the anomalous
Josephson current was obtained, which result means
a realization of $\varphi_0$-state.
Moreover, the direction-dependence of critical current
is found. The critical current indicates cusps at
the local minima. The position of cusps also
depends on the current direction,
in accordance with the experiment.~\cite{private}
Between the cusps for positive and negative
current direction, the transition from
$\varphi_0 \approx 0$ to $\approx \pi$ takes place.

In this paper, we study the anomalous Josephson effect
numerically using a tight-binding model for the nanowire
in the case of short junction.
The purposes are to confirm our previous simple model
and to elucidate the key ingredients of anomalous Josephson effect.
First, we consider the case without SO interaction.
The Andreev levels are invariant against
the $\varphi$-inversion, $E_n (-\varphi ) = E_n (\varphi )$.
As a result, the current satisfies $I(-\varphi) =-I(\varphi)$
and hence no anomalous current is found.
The critical current oscillates as a function of
magnetic field accompanying the $0$-$\pi$ transition,
which is characterized by a single parameter $\theta_B$
even for $N>1$.
Next, we investigate the Josephson effect in
the presence of SO interaction and Zeeman effect.
The relation of $I(-\varphi) =-I(\varphi)$ is
broken. As a result, the anomalous Josephson current
and the direction dependence of critical current
are obtained, which are qualitatively the same as
those of single scatterer model.
We stress that the spin-dependent channel
mixing due to SO interaction plays an important role on
the anomalous Josephson effect.

The organization of this paper is as follows.
In Sec.\ II, we explain our model for the S/NW/S Josephson
junction and calculation method of the Andreev levels and
Josephson current. Numerical results are given in Sec.\ III.
The last section (Sec.\ IV) is devoted to the conclusions
and discussion.

\section{MODEL AND CALCULATION}
\label{sec:model}

\begin{figure}
\includegraphics[width=70mm]{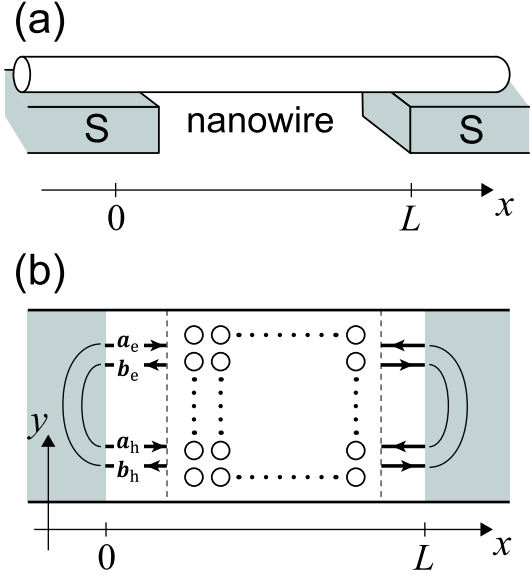}
\caption{
Our model for a quasi-one-dimensional semiconductor
nanowire connected to two superconductors.
The nanowire is along the $x$ axis.
(a) Schematic view of the model. The pair potential is
induced in the nanowire by the proximity effect,
$\Delta (x) = \Delta_0 e^{i\varphi_{\rm L}}$ at $x<0$
and $\Delta_0 e^{i\varphi_{\rm R}}$ at $L<x$, whereas
$\Delta (x) =0$ at $0<x<L$. Several impurities
are present in the nanowire. The spin-orbit interaction and
Zeeman effect are taken into account only in the normal region.
(b) The tight-binding model gives the scattering matrix
$\hat{S}_{\rm e}$ ($\hat{S}_{\rm h}$) for electrons (holes),
which connects incoming $\bm{a}_{\rm e}$ ($\bm{a}_{\rm h}$)
and outgoing electrons $\bm{b}_{\rm e}$ (holes $\bm{b}_{\rm h}$).
At $x=0$ and $L$, the electron $\bm{b}_{\rm e}$ is
reflected as the hole $\bm{a}_{\rm h}$ by the Andreev
reflection, whereas $\bm{b}_{\rm h}$ is reflected as $\bm{a}_{\rm e}$.
}
\label{fig1:Model}
\end{figure}

In this section, we explain our model depicted in Fig.\ \ref{fig1:Model}.
We introduce the Bogoliubov-de Gennes (BdG) equation to obtain
the Andreev levels. The formulation of solving BdG equation
is given in terms of scattering matrix.~\cite{Beenakker1}
We apply the tight-binding model to the normal region,
where the scattering matrices of conduction electrons and
holes are numerically calculated.~\cite{Datta,Ando}

\subsection{Formulation}
\label{sec:formulation}

We consider a semiconductor nanowire along the $x$ axis
connected to two superconductors at $x<0$ and $x>L$,
as shown in Fig.\ \ref{fig1:Model}(a).
The superconducting pair potential is penetrated into
the nanowire by the proximity effect, whereas there is
no pair potential in the normal region at $0<x<L$.
The SO interaction and Zeeman effect in a magnetic field
are taken into account only in the normal region.
Since InSb has a large g-factor, a large Zeeman energy is
obtained for weak magnetic field, which does not break
the superconductivity.

The Andreev bound states are formed in the normal region.
The BdG equation to describe the Andreev bound
states is written as~\cite{Blonder,NazarovBlanter}
\begin{equation}
\left( \begin{array}{cc}
H - E_{\rm F} & \hat{\Delta} \\
\hat{\Delta}^\dagger & -(H^* - E_{\rm F})
\end{array} \right)
\left( \begin{array}{c}
\bm{\psi}_{\rm e} \\
\bm{\psi}_{\rm h}
\end{array} \right)
= E \left( \begin{array}{c}
\bm{\psi}_{\rm e} \\
\bm{\psi}_{\rm h}
\end{array} \right),
\label{eq:BdG}
\end{equation}
where $\bm{\psi}_{\rm e} = (\psi_{{\rm e} +}, \psi_{{\rm e} -} )^{\rm T}$
and $\bm{\psi}_{\rm h} = (\psi_{{\rm h} +}, \psi_{{\rm h} -} )^{\rm T}$
are the spinors for electron and hole, respectively.
The energy $E$ is measured from the Fermi level $E_{\rm F}$.
The diagonal element is the single-electron Hamiltonian
$H = H_0 + H_{\rm SO} + H_{\rm Z}$ with
$H_0 = \bm{p}^2 /(2m^*) + V_{\rm conf} (y,z) + V_{\rm imp}$,
SO interaction $H_{\rm SO}$, and Zeeman effect
$H_{\rm Z} = g \mu_{\rm B} {\bm B} \cdot \hat{\bm{\sigma}}/2$,
using effective mass $m^*$, $g$-factor $g$ 
($< 0$ for InSb), Bohr magneton $\mu_{\rm B}$,
and Pauli matrices $\hat{\bm{\sigma}}$. $H_{\rm SO}$ and
$H_{\rm Z}$ are taken into account only at $0<x<L$.
$V_{\rm conf}$ describes the confining potential of nanowire.
$V_{\rm imp}$ represents the potential due to the impurities.
$\hat{\Delta}$ is the pair potential in the spinor space
\begin{equation}
\hat{\Delta} = \Delta (x) \hat{g} = \Delta (x)
\left( \begin{array}{cc}
 & -1 \\
 1 & 
\end{array} \right),
\label{eq:Deltamatrix}
\end{equation}
where $\hat{g}=-i \hat{\sigma}_y$.~\cite{com1}
For simplicity, we assume that the absolute values of
pair potential in left and right superconducting regions
are equal to each other,
$\Delta (x) = \Delta_0 e^{i\varphi_{\rm L}}$ at $x<0$
and $\Delta_0 e^{i\varphi_{\rm R}}$ at $L<x$.
In the normal region at $0<x<L$, $\Delta (x) =0$.
The phase difference between two superconductors is
defined as $\varphi \equiv \varphi_{\rm L} - \varphi_{\rm R}$.
We consider a short junction, where $L \ll \xi$.
No potential barrier is assumed at the boundaries between
the normal and superconducting regions.
The Zeeman energy $E_{\rm Z} \equiv |g \mu_B B|/2$ and
the pair potential $\Delta_0$ are much smaller than
the Fermi energy $E_{\rm F}$.

The solution of BdG equation gives the Andreev levels $E_n$
($|E_n|<\Delta_0$) as a function of phase difference $\varphi$.
When the BdG equation has an eigenenergy $E_n$ with
eigenvector $(\bm{\psi}_{{\rm e},n}, \bm{\psi}_{{\rm h},n})^{\rm T}$,
$-E_n$ is also an eigenenergy of the equation with
$(\bm{\psi}_{{\rm h},n}^*, \bm{\psi}_{{\rm e},n}^*)^{\rm T}$.
In short junctions, the number of Andreev levels is given by
$4N$; $2N$ positive levels and $2N$ negative ones when
the number of channels is $N$ ($2N$ if the spin degree of
freedom is included). The ground-state energy of the junction
is given by
\begin{equation}
E_{\rm gs} (\varphi) =-\frac{1}{2}
{\sum_n}^\prime E_n (\varphi),
\label{eq:Egs}
\end{equation}
where the summation is taken over all the positive Andreev levels,
$E_n (\varphi)>0$. The contribution from continuous levels
($|E| >\Delta_0$) is disregarded in Eq.\ (\ref{eq:Egs}), which
are independent of $\varphi$ in the short junctions.~\cite{Beenakker1}
At zero temperature, the supercurrent is calculated as
\begin{equation}
I (\varphi) =
\frac{2e}{\hbar} \frac{d E_{\rm gs}}{d \varphi}
=- \frac{e}{\hbar} {\sum_n}^{\prime} \frac{d E_n}{d \varphi}.
\label{eq:JC}
\end{equation}
The current is a periodic function for $-\pi < \varphi \le \pi$.
The maximum (absolute value of minimum) of $I (\varphi)$
yields the critical current $I_{{\rm c},+}$
($I_{{\rm c},-}$) in the positive (negative) direction.

The symmetry of BdG equation should be noted here.
We denote the matrix on the left side of Eq.\ (\ref{eq:BdG}) by
$\mathcal{H} (\varphi)$. In the absence of Zeeman effect,
$\mathcal{T} \mathcal{H} (\varphi) \mathcal{T}^{-1}=
\mathcal{H} (-\varphi)$ with the time-reversal operator
$\mathcal{T}=-i \hat{\sigma}_y K$ for spin-1/2 particles.
$K$ is the operator to form a complex conjugate; $Kf=f^*$.
If $\mathcal{H} (\varphi)$ has an eigenenergy $E_n$ with eigenvector
$(\bm{\psi}_{{\rm e},n}, \bm{\psi}_{{\rm h},n})^{\rm T}$,
$\mathcal{H} (-\varphi)$ has an eigenenergy $E_n$ with eigenvector
$\mathcal{T} (\bm{\psi}_{{\rm e},n}, \bm{\psi}_{{\rm h},n})^{\rm T}$.
Thus the Andreev levels satisfy the relation of
$E_n(\varphi)=E_n(-\varphi)$. In the absence of SO interaction,
$K \mathcal{H} (\varphi) K^{-1}=\mathcal{H} (-\varphi)$.
Then we derive that $E_n(\varphi)=E_n(-\varphi)$ in the same way.
The relation does not always hold in the presence of both
SO interaction and magnetic field.

\subsection{Scattering Matrix Approach}
\label{sec:SMA}

The BdG equation in Eq.\ (\ref{eq:BdG}) can be written in
terms of the scattering matrix.~\cite{Beenakker1}

In the normal region with SO interaction and Zeeman effect,
the quantum transport of electrons (holes) is described by
the scattering matrix $S_{\rm e}$ ($S_{\rm h}$).
The scattering matrix $S_{\rm p}$ (${\rm p} = {\rm e,h}$)
connects the amplitudes of incoming waves of
$N$ conduction channels with spin $\sigma =\pm$,
$(\bm{a}_{\rm pL}, \bm{a}_{\rm pR})^{\rm T}$, and
those of outgoing waves, $(\bm{b}_{\rm pL}, \bm{b}_{\rm pR})^{\rm T}$,
as shown in Fig.\ \ref{fig1:Model}(b),
\begin{equation}
\left( \begin{array}{c}
\bm{b}_{\rm pL} \\
\bm{b}_{\rm pR}
\end{array} \right)
= \hat{S}_{\rm p}
\left( \begin{array}{c}
\bm{a}_{\rm pL} \\
\bm{a}_{\rm pR}
\end{array} \right).
\label{eq:Smatrix}
\end{equation}
$\hat{S}_{\rm e}$ and $\hat{S}_{\rm h}$ are
$4N \times 4N$ matrices and related to each other by
$\hat{S}_{\rm e}(E)=\hat{S}_{\rm h}^*(-E)$.
On the assumption that they are independent of
energy $E$ for $|E|<\Delta_0$, and thus
$\hat{S}_{\rm e}=\hat{S}_{\rm h}^*$.
We denote $\hat{S}_{\rm e} = \hat{S}$ and
$\hat{S}_{\rm h} = \hat{S}^*$.
$\hat{S}$ is conventionally written by reflection
and transmission matrices:
\begin{equation}
\hat{S} = \left( \begin{array}{cc}
 \hat{r}_{\rm L}   & \hat{t}_{\rm LR} \\
 \hat{t}_{\rm RL} & \hat{r}_{\rm R}
\end{array} \right).
\end{equation}
The scattering matrix is unitary,
$\hat{S}^\dagger \hat{S} =\hat{1}$.
Moreover,
$\hat{r}_{\rm L}^{\rm T}=\hat{g}^\dagger \hat{r}_{\rm L} \hat{g}$,
$\hat{r}_{\rm R}^{\rm T}=\hat{g}^\dagger \hat{r}_{\rm R} \hat{g}$,
and
$\hat{t}_{\rm RL}^{\rm T}=\hat{g}^\dagger \hat{t}_{\rm LR}  \hat{g}$
are satisfied if the time reversal symmetry is kept.

The Andreev reflection at $x=0$ and $L$ is
also described in terms of scattering matrix
$\hat{r}_{\rm he}$ for the conversion from
electron to hole and $\hat{r}_{\rm eh}$ for
that from hole to electron. When an electron with
spin $\sigma$ is reflected into a hole with $-\sigma$,
it is written as~\cite{Beenakker1}
\begin{equation}
\left( \begin{array}{c}
\bm{a}_{\rm hL} \\
\bm{a}_{\rm hR}
\end{array} \right)
= \hat{r}_{\rm he}
\left( \begin{array}{c}
\bm{b}_{\rm eL} \\
\bm{b}_{\rm eR}
\end{array} \right),
\end{equation}
where
\begin{equation}
\hat{r}_{\rm he} = e^{-i \alpha_{\rm A}}
\left( \begin{array}{cc}
 e^{-i \varphi_{\rm L}} \hat{1} \otimes \hat{g} & \\
 & e^{-i \varphi_{\rm R}} \hat{1} \otimes \hat{g}
\end{array} \right)
\label{eq:AR1}
\end{equation}
with $\alpha_{\rm A} \equiv \arccos(E /\Delta_0)$.
When a hole is reflected to an electron, it is
\begin{equation}
\left( \begin{array}{c}
\bm{a}_{\rm eL} \\
\bm{a}_{\rm eR}
\end{array} \right)
= \hat{r}_{\rm eh}
\left( \begin{array}{c}
\bm{b}_{\rm hL} \\
\bm{b}_{\rm hR}
\end{array} \right)
\end{equation}
with
\begin{equation}
\hat{r}_{\rm eh} = e^{-i \alpha_{\rm A}}
\left( \begin{array}{cc}
 e^{i \varphi_{\rm L}} \hat{1} \otimes \hat{g}^\dagger & \\
 & e^{i \varphi_{\rm R}} \hat{1} \otimes \hat{g}^\dagger
\end{array} \right).
\label{eq:AR2}
\end{equation}
We assume that the channel is conserved at
the Andreev reflection in the case of $N \ge 2$.
The normal reflection can be neglected in our case
without potential barriers at the boundaries.~\cite{Andreev}

The product of scattering matrices gives
an equation for $(\bm{a}_{\rm eL}, \bm{a}_{\rm eR})^{\rm T}$.
The Andreev levels $E_n (\varphi)$ are
calculated from this product as,~\cite{Beenakker1}
\begin{equation}
\det \left(\hat{1} - \hat{r}_{\rm eh}
\hat{S}^* \hat{r}_{\rm he}\hat{S} \right) =0,
\label{eq:determinant}
\end{equation}
which is equivalent with the BdG equation in Eq.\ (\ref{eq:BdG}).
In the absence of magnetic field, Eq.\ (\ref{eq:determinant}) is
simply reduced to~\cite{Beenakker1}
\begin{equation}
\det \left[ 1 - \left( \frac{E}{\Delta_0} \right)^2
- \hat{t}_{\rm LR}^\dagger \hat{t}_{\rm LR} \sin^2
\left( \frac{\varphi }{2} \right) \right] = 0.
\label{eq:detnoB}
\end{equation}
In this case, the Andreev levels are represented by
the transmission eigenvalues of
$\hat{t}_{\rm LR}^\dagger \hat{t}_{\rm LR}$.
They are two-fold degenerate reflecting
the Kramers' degeneracy at $\varphi =0$.
The Andreev levels $E_n(\varphi)$ are not split by
finite $\varphi$ in spite of the broken time reversal symmetry.

\subsection{Tight-Binding Model}
\label{sec:TBM}

To obtain the scattering matrix $\hat{S}$,
we describe the normal region by a tight-binding
model of square lattice model in two-dimensional
space ($xy$ plane),~\cite{Datta} as schematically
shown in Fig.\ \ref{fig1:Model}(b).
We consider a quasi-one-dimensional nanowire
along the $x$ axis with width $W$ in the $y$ direction.
The length of normal region is $L$.
We assume hard-wall potentials at $y=0$ and $W$.
The Rashba-type SO interaction in Eq.\ (\ref{eq:Rashba})
and the Zeeman effect are considered in the normal region.
The Rashba interaction specifies the direction of
spin quantization axis. In the experiments,
the nanowire is not two-dimensional or
the SO interaction may not be Rashba one.
However, our model is general to represent
a single or few conduction channels in the nanowire
and to consider the spin mixing among channels
by the SO interaction.
In the following, the magnetic field is applied in
the $y$ direction, which is almost parallel to
the spin quantization axis due to the Rashba interaction
for the channels. The channel is split
upward and downward by the Zeeman effect.
The orbital magnetization is neglected.

On the tight-binding model, the Hamiltonian
$H = H_0 + H_{\rm SO} + H_{\rm Z}$ is written as
\begin{eqnarray}
H &=& t \sum_{j,l}
\bm{c}_{j,l}^\dagger
\left\{ (4 + v_{j,l}) \hat{1}
+ \bm{b} \cdot \hat{\bm{\sigma}} \right\}
\bm{c}_{j,l} \nonumber \\
&& -t \sum_{j,l} \left(
\bm{c}_{j,l}^\dagger \hat{T}_{j,l;j+1,l} \bm{c}_{j+1,l}
+ \bm{c}_{j,l}^\dagger \hat{T}_{j,l;j,l+1} \bm{c}_{j,l+1}
+ \text{H.\ c.} \right),
\label{eq:tbHamiltonian}
\end{eqnarray}
where $\bm{c}_{j,l} \equiv (c_{j,l;+} ,c_{j,l;-})^{\rm T}$ and
$c_{j,l;\sigma}$ is annihilation operator of an electron at
site $(j,l)$ with spin $\sigma$. $t \equiv \hbar^2 /(2m^* a^2)$
is a transfer integral with a lattice constant $a$.
Here, $j$ ($=0,1,\cdots ,N_x ,N_x +1$) and $l$
($=1,2, \cdots, N_y$) denote site labels in
the $x$ and $y$ directions, respectively.
The length is $L=N_x a$ and the width is $W=(N_y +1)a$.
At the sites of $j=0,N_x +1$, the SO interaction and
Zeeman effect is absent. $v_{j,l} \equiv V_{j,l}/t$ is
a dimensionless on-site potential by impurities.
$\hat{1}$ is the unit matrix in the spinor space.
$\bm{b} \equiv g\mu_{\rm B} \bm{B}/(2t)$
indicates a magnetic field.
The transfer term in the $x$ direction is given by
\begin{equation}
\hat{T}_{j,l;j+1,l} = \hat{1} -i k_\alpha a \hat{\sigma}_y,
\label{eq:xhopping}
\end{equation}
whereas that in the $y$ direction is
\begin{equation}
\hat{T}_{j,l;j,l+1} = \hat{1} +i k_\alpha a \hat{\sigma}_x.
\label{eq:yhopping}
\end{equation}
Here, $k_\alpha = m^* \alpha /\hbar^2$ denotes
the strength of Rashba interaction.
In this model, the reflection and transmission matrices
are calculated by using the recursive Green's function
method (see Appendix A).~\cite{Ando}

We set the Fermi wavelength $\lambda_{\rm F}$
as a parameter, which gives the Fermi energy by
$E_{\rm F} = 2t - 2t \cos (k_{\rm F} a)$
with $k_{\rm F} = 2\pi /\lambda_{\rm F}$.~\cite{comSO}
In an ideal quantum wire with width $W$, the dispersion
relation for channel $n$ is given by
\begin{equation}
E_n (k)= 4t  - 2t\cos (k a) - 2t \cos \left( \frac{\pi a}{W} n \right).
\label{eq:dispersion}
\end{equation}
The conduction channels satisfy $E_n (k=0) <E_{\rm F}$.
Then, the velocity of channel $n$ at the Fermi energy is
\begin{equation}
v_{{\rm F},n} = \frac{2ta}{\hbar}
\sqrt{ 1 - \left\{ 2 - \cos \left( \frac{\pi a}{W} n \right)
- \frac{E_{\rm F}}{2t} \right\}^2 }.
\label{eq:velocity}
\end{equation}

We consider a nanowire with width $W=60\, \mathrm{nm}$.
The distance between left and right superconductors is
$L=1000\, \mathrm{nm}$. We set $N_x =11$ and $N_y =200$.
The number of conduction channel is changed by tuning
the Fermi wavenumber $\lambda_{\rm F}$.
In the following, we calculate for three cases:
$\lambda_{\rm F} = 90\, \mathrm{nm}$ for single channel
($N=1$), $50\, \mathrm{nm}$ for $N=2$,
and $25\, \mathrm{nm}$ for $N=4$.
A parameter of Rashba interaction is $k_\alpha /k_{\rm F} =0.15$.
The on-site random potential by impurities is
uniformly distributed in $-W_0 /2 < V_{j,l} < W_0 /2$.
The mean free path $l_0$ is estimated as~\cite{Ando}
\begin{equation}
l_0 =\frac{6\lambda_{\rm F}^3}{\pi^3 a^2}
\left( \frac{E_{\rm F}^\prime}{W_0} \right)^2.
\end{equation}
Here we use the modified Fermi energy
$E_{\rm F}^\prime = E_{\rm F} - E_1 (0)$ in
an one-dimensional quantum wire.

\section{NUMERICAL RESULTS}

In this section, we present calculated results of
Andreev levels and Josephson currents.
First, we discuss the case without SO interaction.
The critical current oscillates as a function of
magnetic field and the $0$-$\pi$ transition is clearly found.
Next, we consider the anomalous Josephson effect
induced by the SO interaction.

The magnetic field is applied in the $y$ direction. 
We introduce a parameter of magnetic field,
$\theta_B = |g|\mu_{\rm B}BL/(\hbar \bar{v}_{\rm F})$, where
$\bar{v}_{\rm F} \equiv \left\{ (1/N)
\sum_{n=1}^N (1/v_{{\rm F},n}) \right\}^{-1}$
is the inversion average of velocity $v_{{\rm F},n}$
in Eq.\ (\ref{eq:velocity}).
The Zeeman effect splits the dispersion relation for
spin $\sigma =\pm$. The wavenumber becomes
$k_{{\rm F},n,\pm} \simeq k_{{\rm F},n}
\pm |g|\mu_{\rm B}B/(2\hbar v_{{\rm F},n})$
for $k>0$ and
$k_{{\rm F},n,\pm} \simeq - k_{{\rm F},n}
\mp |g|\mu_{\rm B}B/(2\hbar v_{{\rm F},n})$
for $k<0$.
For the propagation of electron with spin $\sigma =\pm$
and hole with $\sigma =\mp$ in the normal region,
the shift of phase due to the Zeeman effect is
$\pm |g|\mu_{\rm B}BL/(\hbar v_{{\rm F},n})$.
Therefore, $\theta_B$ means the channel-average of
spin-dependent phase shift of electron and hole forming
the Andreev bound states.

\subsection{Absence of spin-orbit interaction}
\label{sec:noSO}

\subsubsection{Single conduction channel}
\label{sec:noSO1}

\begin{figure}
\includegraphics[width=80mm]{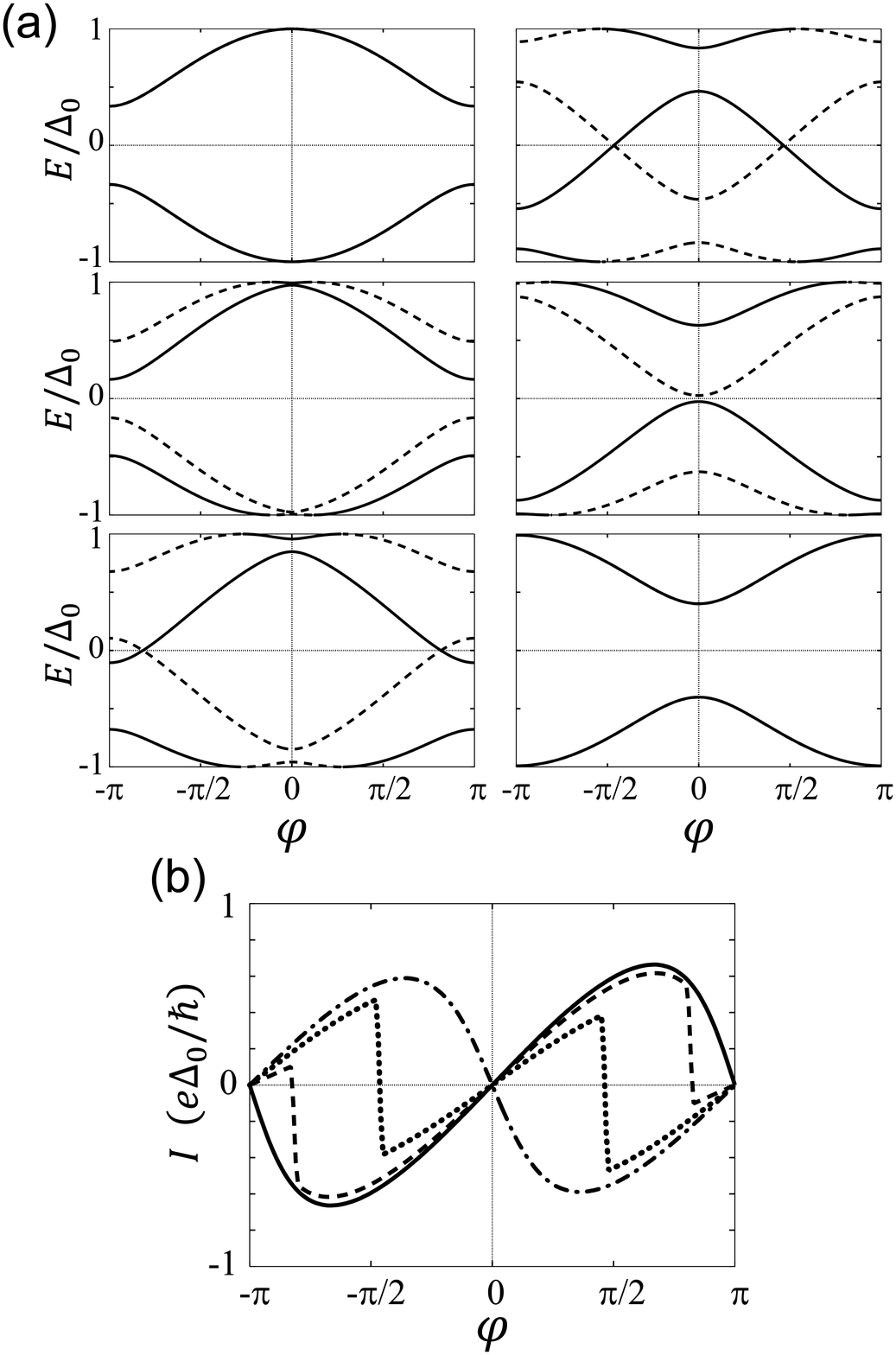}
\caption{
Calculated results for a sample when $N=1$ and $l_0/L=1$.
The SO interaction is absent.
(a) Andreev levels $E_n$ as a function of phase difference
$\varphi$ between two superconductors. Solid and broken
lines indicate $E_{\uparrow i \pm}$ and $E_{\downarrow i \pm}$,
respectively. The magnetic field is $\theta_B = 0$ (left upper),
$0.1\pi$ (left middle), $0.27\pi$ (left bottom), $0.53\pi$ (right upper),
$0.79\pi$ (right middle), and $\pi$ (right bottom).
At $B = 0$, two lines are overlapped to each other,
reflecting the Kramers' degeneracy.
(b) Josephson current $I(\varphi)$ through the nanowire when
$\theta_B=0$ (solid), $0.27\pi$ (broken), $0.53\pi$ (dotted),
and $\pi$ (dotted broken lines).
}
\label{fig2:N1AL}
\end{figure}

\begin{figure}
\includegraphics[width=65mm]{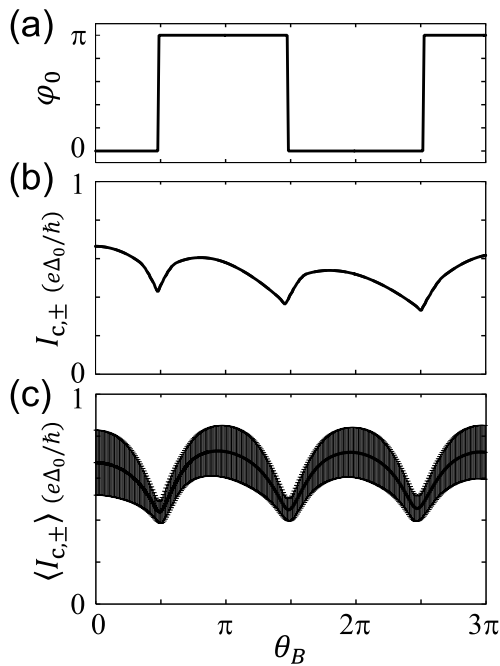}
\caption{
Calculated results for $N=1$ and $l_0/L=1$.
The SO interaction is absent.
(a) Phase difference $\varphi_0$ at the minimum of
ground-state energy as a function of magnetic field, $\theta_B$.
(b) Critical current $I_{{\rm c},\pm}$. The current in the positive
direction $I_{{\rm c}, +}$ is identical with that in the
negative direction $I_{{\rm c}, -}$. The sample for
(a) and (b) is same as that in Fig.\ \ref{fig2:N1AL}.
(c) Average of critical current, $\langle I_{{\rm c},\pm} \rangle$,
with the average of fluctuation,
$\sqrt{\langle [\Delta I_{{\rm c},\pm}]^2 \rangle}$,
as error bars, where $\Delta I_{{\rm c},\pm} \equiv
I_{{\rm c},\pm} - \langle I_{{\rm c},\pm} \rangle$.
The random average is taken for 400 samples.
}
\label{fig3:N1IC}
\end{figure}

First, we consider a sample of nanowire with
a single conduction channel. Figure \ref{fig2:N1AL} shows
the Andreev levels and Josephson currents as
functions of phase difference $\varphi$ between
two superconductors. The magnetic field gradually
increases from left-upper to right-bottom panels.
In the absence of SO interaction, the spin $\sigma =\pm$
is well defined in the direction of magnetic field.
In the case of single conduction channel,
four Andreev levels are found in $|E| \le \Delta_0$.
The levels are denoted as $E_{\uparrow i \pm}$ and
$E_{\downarrow i \pm}$, where the subscript
$\uparrow$ ($\downarrow$) means the state of electron spin
$\sigma=+1$ ($\sigma=-1$) and hole spin
$\sigma=-1$ ($\sigma=+1$). $i=1,2,\cdots$.
The sign $\pm$ corresponds to the positive or
negative energy at $B=0$. We consider three regions
with increasing $\theta_B$. When $B=0$,
the levels are doubly degenerate for any $\varphi$.
The ground-state energy $E_{\rm gs}$ becomes
minimal at $\varphi =0$, which corresponds to
the $0$-state [Fig.\ \ref{fig3:N1IC}(a)].
The levels are split like the Zeeman splitting in
the presence of magnetic field.
For a weak magnetic field, $E_{\rm gs}$ is still
minimal at $\varphi =0$ [region (I)].
As the magnetic field is increased, the level crossing
at $E=0$ is observed, which corresponds to region (II).
The crossing points move from $\varphi = \pm \pi$ to
$0$ with increase of $\theta_B$. When $\theta_B \approx \pi$,
no level crossing takes place in region (III).
In this region, $E_{\rm gs}$ is minimal at
$\varphi =\pi$ ($\pi$-state).
The transition of $0$-state to $\pi$-state
takes place suddenly around $\theta_B = \pi/2$,
as shown in Fig.\ \ref{fig3:N1IC}(a).
This is called the $0$-$\pi$ transition.
With increase of magnetic field, some levels go to $|E|>\Delta_0$.
At the same time, another levels come into $|E| \le \Delta_0$.
Therefore, the number of Andreev levels in $|E| \le \Delta_0$ is fixed.

The Josephson current is calculated from the sum of positive
Andreev levels in Eq.\ (\ref{eq:JC}). In Fig.\ \ref{fig2:N1AL}(a),
the Andreev levels $E_n (\varphi)$ are invariant against
the inversion of $\varphi$, $E_n (-\varphi) = E_n (\varphi)$.
As a result, the Josephson current satisfies
$I (-\varphi) = -I (\varphi)$ in Fig.\ \ref{fig2:N1AL}(b).
When $\theta_B =0$, the current $I (\varphi)$ is similar to
$\sin \varphi$, which is a feature of $0$-state.
When the level crossing takes place at $E=0$,
the crossing results in the discontinuity in the current
phase relation. Around $\theta_B = \pi/2$, a saw-tooth
current phase relation is obtained. The discontinuous points
move from $\pm \pi$ to $0$. When $\theta_B \approx \pi$,
the current is roughly $I (\varphi ) \sim -\sin \varphi$,
which is a feature of $\pi$-state.

Figure \ref{fig3:N1IC}(b) shows the critical current as
a function of $\theta_B$. Since $I (-\varphi) = -I (\varphi)$,
the maximum of Josephson current, $I_{{\rm c},+}$,
is identical with the absolute value of minimum of
current, $I_{{\rm c},-}$.
When the magnetic field is stronger, the phase difference
$\varphi_0$ at the minimum of ground-state energy
changes between $0$ and $\pi$ discontinuously at
$\theta_B \approx (2m+1)\pi/2$, where
$m=0,1,2,\cdots$ [Fig.\ \ref{fig3:N1IC}(a)].
The critical current oscillates with cusps around
the $0$-$\pi$ transitions. In Fig.\ \ref{fig3:N1IC}(c),
we plot a random average $\langle I_{{\rm c},\pm} \rangle$ of
the critical current with the fluctuation
$\sqrt{\langle [\Delta I_{{\rm c},\pm}]^2 \rangle}$.
The fluctuation is defined as $\sqrt{\langle [\Delta A]^2 \rangle}$
with $\Delta A \equiv A-\langle A \rangle$.
$\langle I_{{\rm c},\pm} \rangle$ also exhibits the cusps at
$\theta_B \approx (2m+1)\pi/2$, where its fluctuation is
relatively small. When the Fermi energy is tuned,
$\theta_{\rm B}$ is also modified via the velocity
$v_{{\rm F},1}$. However, the cusp is always found at
$\theta_B \approx (2m+1)\pi/2$ (not shown).

\subsubsection{A few conduction channels}
\label{sec:noSO2}

\begin{figure}
\hspace{-10mm}
\includegraphics[width=95mm]{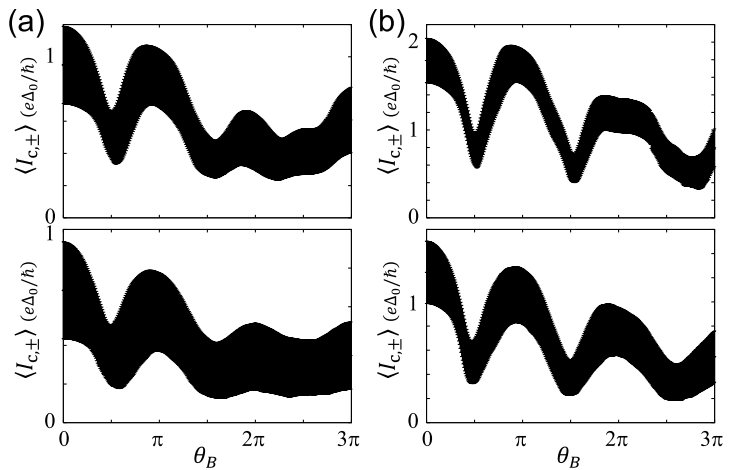}
\caption{
Calculated results of random average of critical current,
$\langle I_{{\rm c},\pm} \rangle$ as a function of magnetic field,
$\theta_B$ when $N=2$ (a) and $4$ (b).
Error bars represent the average of fluctuation,
$\sqrt{\langle [\Delta I_{{\rm c},\pm}]^2 \rangle}$, where
$\Delta I_{{\rm c},\pm} \equiv I_{{\rm c},\pm} - \langle I_{{\rm c},\pm} \rangle$.
The SO interaction is absent. The mean free path is
$l_0 /L =1$ (upper) and $0.5$ (lower panels).
The random average is taken for 400 samples.
}
\label{fig4:N24IC}
\end{figure}

Next, we consider the case of a few conduction channels in
the nanowire. For $N$ conduction channels, $2N$ positive and
$2N$ negative Andreev levels are obtained even if the channels
are mixed with each other by the impurity scattering.
The behavior of Andreev levels in magnetic field is qualitatively
the same as in Fig.\ \ref{fig2:N1AL}(a) except the number of levels.
The Andreev levels keep the relation $E (-\varphi) = E (\varphi)$,
which results in the current $I (-\varphi) = -I (\varphi)$.
The critical current is independent of its current direction.
When the magnetic field is applied, the critical current
oscillates accompanying the $0$-$\pi$ transition around
the local minima of $I_{{\rm c},\pm}$.

Upper and lower panels in Fig.\ \ref{fig4:N24IC}(a)
show the random average of critical current for $N=2$
when the mean free path is $l_0 /L =1$ and $0.5$, respectively.
The average $\langle I_{{\rm c},\pm} \rangle$ of
critical current oscillates as a function of magnetic field
and indicates the first and second local minima at
$\theta_B \approx \pi /2$ and $3\pi /2$, respectively.
The positions of the two local minima are hardly shifted by
the impurity scattering, or the mean free path $l_0$.
$\langle I_{{\rm c},\pm} \rangle$ becomes local minimal
also around $\theta_B = 5\pi /2$ when $l_0 /L =0.5$,
whereas the position of the third local minimum is
shifted from $\theta_B = 5\pi /2$ in the case of $l_0 /L =1$.
Figure \ref{fig4:N24IC}(b) shows $\langle I_{{\rm c},\pm} \rangle$
in the case of $N=4$, where $\langle I_{{\rm c},\pm} \rangle$
is local minimal at $\theta_B \approx \pi /2$ and $3\pi /2$.
For both cases of $N=2$ and $4$, the local minima of
$\langle I_{{\rm c},\pm} \rangle$ tend to be located at
$\theta_B \approx (2m+1)\pi/2$ when the impurity
scattering is stronger. This period is the same as that of $N=1$.

\subsection{Presence of spin-orbit interaction}
\label{sec:SO}

\subsubsection{Anomalous Josephson effect}
\label{sec:SO1}

\begin{figure}
\includegraphics[width=80mm]{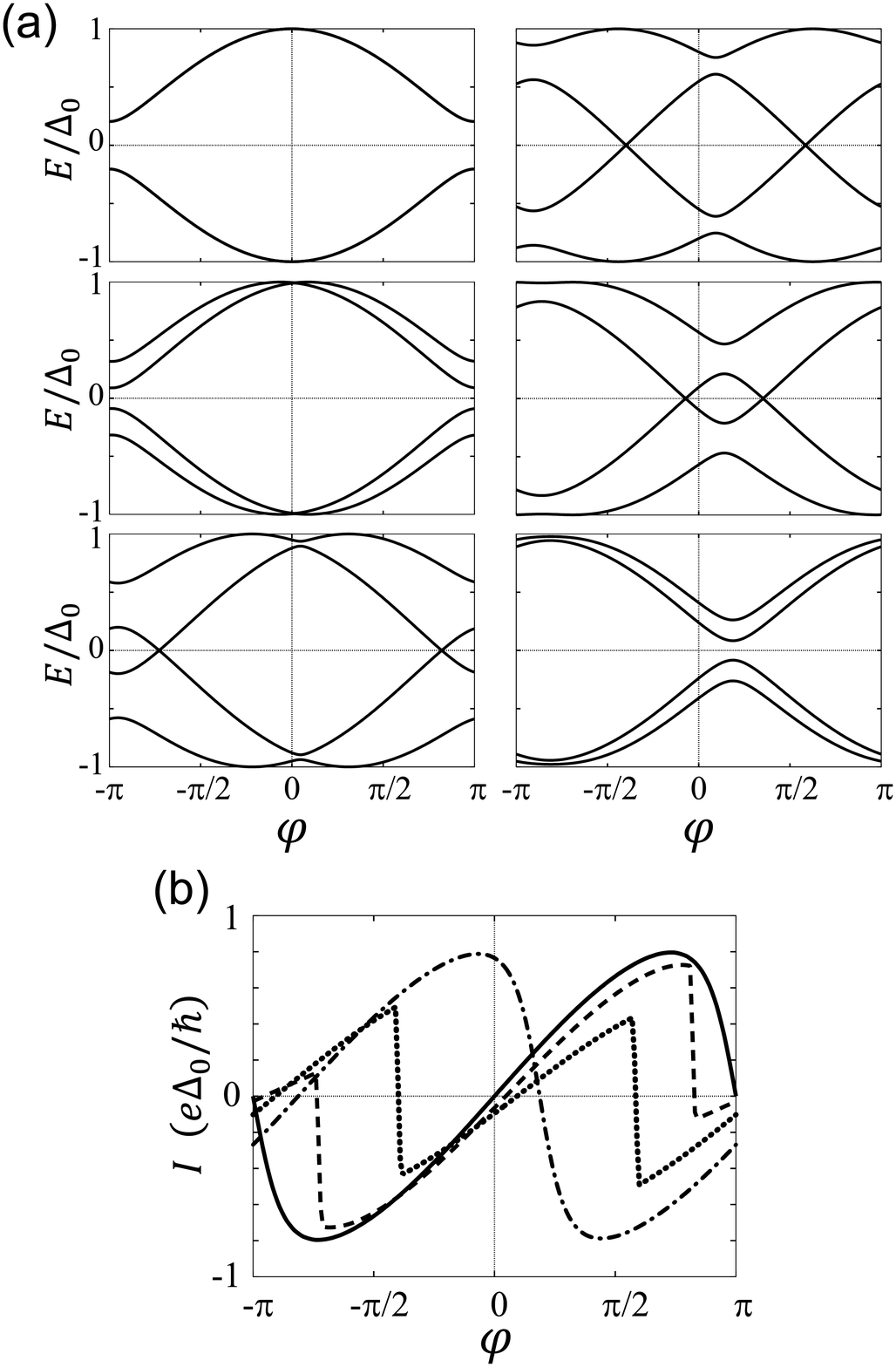}
\caption{
Calculated results for a sample when $N=1$ and $l_0/L=1$.
The SO interaction is $k_\alpha /k_{\rm F} =0.15$.
The magnetic field is applied to the $y$ direction.
(a) Andreev levels $E_n$ as a function of phase difference
$\varphi$ between two superconductors.
The magnetic field is $\theta_B = 0$ (left upper),
$0.1\pi$ (left middle), $0.35\pi$ (left bottom), $0.7\pi$ (right upper),
$1.05\pi$ (right middle), and $1.4\pi$ (right bottom).
At $B = 0$, two lines are overlapped to each other,
reflecting the Kramers' degeneracy.
(b) Josephson current $I(\varphi)$ through the nanowire when
$\theta_B=0$ (solid), $0.35\pi$ (broken), $0.7\pi$ (dotted),
and $1.4\pi$ (dotted broken lines).
}
\label{fig5:N1AJE}
\end{figure}

In this section, we consider the SO interaction in the nanowire.
The SO interaction qualitatively modifies the Andreev levels
in the presence of magnetic field. Figure \ref{fig5:N1AJE}
shows the Andreev level and the Josephson current for
a sample in the case of $N=1$. We assume that
the strength of SO interaction is $k_\alpha /k_{\rm F} =0.15$.
The mean free path is $l_0 /L =1$. In the absence of
magnetic field, the time-reversal symmetry is kept and
the Andreev levels are two-fold degenerate even when
$\varphi \ne 0$ in the case of short junction.~\cite{com2}
The levels satisfy $E (-\varphi) = E (\varphi)$ and
the ground-state energy $E_{\rm gs} (\varphi)$ is
minimum at $\varphi =0$. As the magnetic field is increased,
we find three regions as well as the case without
SO interaction in Sec.\ \ref{sec:noSO1}.
In region (I), for a weak magnetic field, some levels are
positive for any phase difference and the others are
negative although the levels are split by the magnetic field.
With increase of $\theta_B$, the splitting is larger
and the level crossing takes place at $E=0$ in region (II).
This level crossing disappears when the magnetic field is
$\theta_B \approx \pi$ [region (III)].

\begin{figure}
\includegraphics[width=60mm]{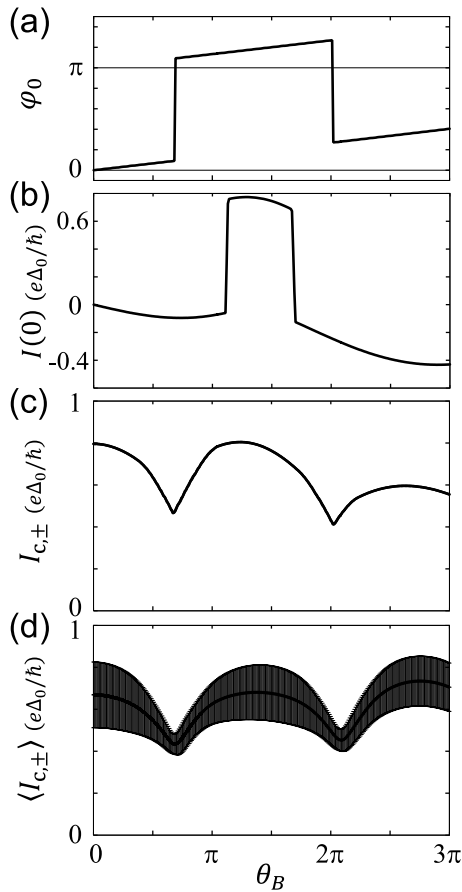}
\caption{
Calculated results for $N=1$ and $l_0/L=1$.
The SO interaction is $k_\alpha /k_{\rm F} =0.15$.
The magnetic field is applied to the $y$ direction.
(a) Phase difference $\varphi_0$ at the minimum of
ground-state energy as a function of magnetic field, $\theta_B$.
(b) Anomalous Josephson current $I (\varphi =0)$.
(c) Critical current $I_{{\rm c},\pm}$.The current in
the positive direction $I_{{\rm c}, +}$ is identical with
that in the negative direction $I_{{\rm c}, -}$.
The sample for (a), (b), and (c) is same as that in
Fig.\ \ref{fig5:N1AJE}.
(d) Average of critical current, $\langle I_{{\rm c},\pm} \rangle$,
with the average of fluctuation,
$\sqrt{\langle [\Delta I_{{\rm c},\pm}]^2 \rangle}$, as error bars,
where $\Delta I_{{\rm c},\pm} \equiv
I_{{\rm c},\pm} - \langle I_{{\rm c},\pm} \rangle$.
The random average is taken for 400 samples.
}
\label{fig6:N1CC}
\end{figure}

The finding of the regions is the same as that without
SO interaction, whereas the invariance of levels against
the $\varphi$-inversion is broken when $\theta_B \ne 0$.
As a result, the phase difference $\varphi_0$ at
the minimum of ground-state energy is deviated from $0$ or $\pi$,
as shown in Fig.\ \ref{fig6:N1CC}(a), and the $\varphi_0$-state
is realized.~\cite{Buzdin2,Reynoso1,Reynoso2}
The phase difference $\varphi_0$ is almost liner to
the magnetic field first, and jumps to $\varphi_0 \approx \pi$
like the $0$-$\pi$ transition. After the transition,
$\varphi_0$ increases gradually with increase of $\theta_B$,
the slope of which is almost the same as that in
the `$0$ like'-state. At $\theta_B \approx 2\pi$,
the `$\pi$ like'-state transits back to
the `$0$ like'-state. This behavior is understood as
the $0$-$\pi$ transition with additional phase shift.

In Fig.\ \ref{fig5:N1AJE}(b), the Josephson current is
calculated from the Andreev levels in Fig.\ \ref{fig5:N1AJE}(a).
When $B=0$, the current satisfies $I (-\varphi) = -I (\varphi)$,
whereas this relation is broken in the magnetic field.
With increasing magnetic field, the discontinuous points of
current corresponding to the level crossings at $E=0$
are found. $I(\varphi )$ indicates a saw-tooth behavior
when $\theta_B \approx 0.7\pi$. The discontinuous points
vanish at $\theta_B = 1.4\pi$.
Compared with those in Fig.\ \ref{fig2:N1AL}(b),
the current phase relation is gradually moved to right in
the panel as the magnetic field is increased. As a result,
a finite supercurrent at $\varphi =0$ (anomalous Josephson
current) is obtained [Fig.\ \ref{fig6:N1CC}(b)].
For a weak magnetic field, the anomalous current is
negative since the shift of current phase relation is positive
($\varphi_0 >0$). In the `$\pi$ like'-state, the positive anomalous
supercurrent is obtained, where $|I(\varphi =0)|$ is enlarged
up to $0.7 e\Delta_0/\hbar$.

Figure \ref{fig6:N1CC}(c) indicates the critical current
$I_{{\rm c},\pm}$ as a function of $\theta_B$.
Although the relation $I (-\varphi) = -I (\varphi)$
does not hold, the critical currents for positive and negative
directions are identical with each other in the case of $N=1$.
The critical current oscillates with cusps at the local minima.
The distance of cusps is longer than that in Fig.\ \ref{fig3:N1IC}(b),
which is caused by the modification of Fermi velocity due to
the SO interaction. The random average
$\langle I_{{\rm c},\pm} \rangle$ of critical current indicates
local minima at $\theta_B \approx 0.7\pi$ and $\approx 2.1\pi$,
as shown in Fig.\ \ref{fig6:N1CC}(d).
The fluctuation is also small around the minima of
$\langle I_{{\rm c},\pm} \rangle$, where the order of fluctuation
is $0.1e\Delta_0 /\hbar$. These behaviors are qualitatively
the same as those in Fig.\ \ref{fig3:N1IC}(c).

\subsubsection{Direction-dependent critical current}

Here, we consider the case of four conduction channels
and demonstrate that the critical current depends on
the current direction.

\begin{figure}
\includegraphics[width=80mm]{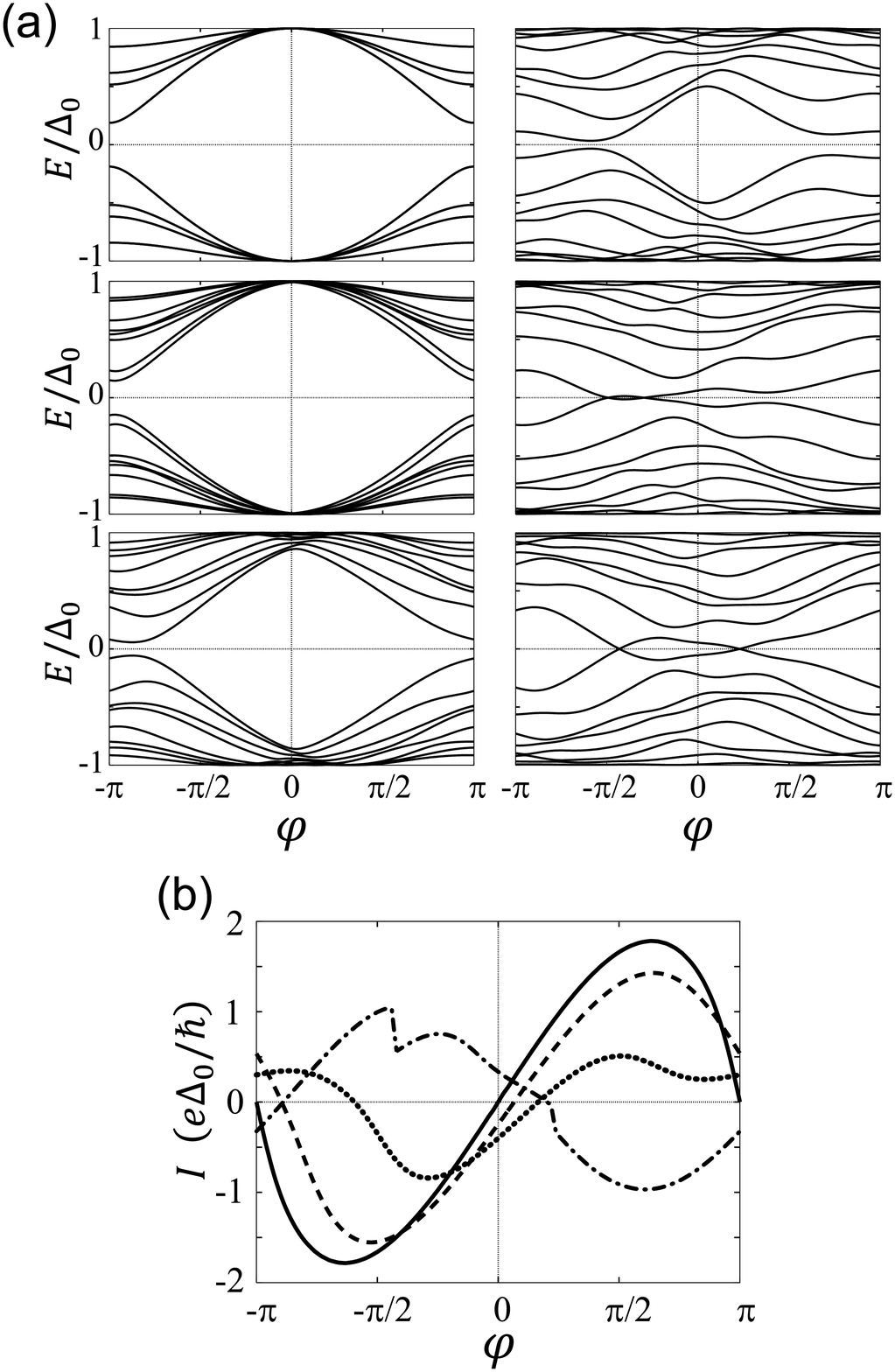}
\caption{
Calculated results for a sample when $N=4$ and $l_0/L=1$.
The SO interaction is $k_\alpha /k_{\rm F} =0.15$.
The magnetic field is applied to the $y$ direction.
(a) Andreev levels $E_n$ as a function of phase difference
$\varphi$ between two superconductors.
The magnetic field is $\theta_B = 0$ (left upper),
$0.1\pi$ (left middle), $0.4\pi$ (left bottom), $0.8\pi$ (right upper),
$1.2\pi$ (right middle), and $1.6\pi$ (right bottom).
At $B = 0$, two lines are overlapped to each other,
reflecting the Kramers' degeneracy.
(b) Josephson current $I(\varphi)$ through the nanowire when
$\theta_B=0$ (solid), $0.4\pi$ (broken), $0.8\pi$ (dotted),
and $1.6\pi$ (dotted broken lines).
}
\label{fig7:N4AJE}
\end{figure}

Figures \ref{fig7:N4AJE}(a) and (b) exhibit the Andreev levels and
the Josephson current, respectively, as functions of $\varphi$
when the magnetic field increases. In each panel of Fig.\ \ref{fig7:N4AJE}(a),
eight positive and eight negative levels are obtained.
In the absence of magnetic field, the levels are doubly
degenerate and invariant against the inversion of $\varphi$.
Four channels are mixed by the impurity scattering and
SO interaction, which contribute to form the Andreev bound states.
In the presence of magnetic field, the Zeeman effect splits
these mixed levels. Then, the Andreev levels become
complicated function of $\varphi$ and the symmetry of
$E (-\varphi) = E (\varphi)$ is broken. As a result,
the Josephson current indicates $I (-\varphi) \ne -I (\varphi)$
in Fig.\ \ref{fig7:N4AJE}(b), where not only the anomalous current,
$I (\varphi = 0) \ne 0$, but also the difference between
maximum and absolute value of minimum currents,
$I_{{\rm c},+} \ne I_{{\rm c},-}$, are obtained.

\begin{figure}
\includegraphics[width=60mm]{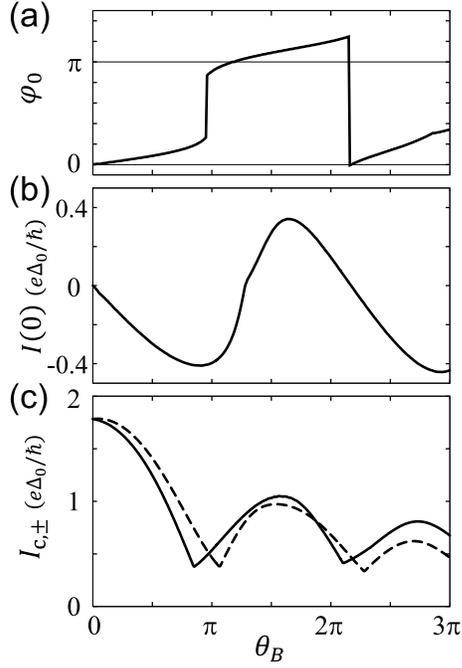}
\caption{
Calculated results for $N=4$ and $l_0/L=1$.
The SO interaction is $k_\alpha /k_{\rm F} =0.15$.
The magnetic field is applied to the $y$ direction.
The sample is same as that in Fig.\ \ref{fig7:N4AJE}.
(a) Phase difference $\varphi_0$ at the minimum of
ground-state energy as a function of magnetic field, $\theta_B$.
(b) Anomalous Josephson current $I (\varphi =0)$.
(c) Critical current in the positive $I_{{\rm c}, +}$ (solid) and
in the negative direction $I_{{\rm c}, -}$ (broken lines).
}
\label{fig8:N4CC}
\end{figure}

We mention the three regions corresponding to region (I), (II),
and (III) described in previous sections. The Josephson current
roughly indicates $I (\varphi) \sim \sin \varphi$ at $\theta_B =0$.
This is the feature of $0$-state in region (I) at $\theta_B \sim 0$.
At $\theta_B =1.6\pi$, the current becomes roughly
$I (\varphi) \sim -\sin \varphi$, which corresponds to
the feature of $\pi$-state in region (III) although the crossing of
Andreev levels at $E=0$ is found. The phase difference
$\varphi_0$ at the minimum of $E_{\rm gs}$ also indicates 
the feature of these two regions: $\varphi_0 \approx 0$ at
$\theta_B \sim 0$ and $\varphi_0 \approx \pi$ at
$\theta_B \sim 1.5\pi$, as shown in Fig.\ \ref{fig8:N4CC}(a).
As the magnetic field is increased, $\varphi_0$ monotonically
increases until $\theta_B \simeq 2.1\pi$.
The `$0$-$\pi$ like' transition occurs at $\theta_B \approx \pi$.
The boundaries between these regions and region (II)
are unclear since the SO interaction tends to avoid
the level crossing at $E=0$. When $\theta_B$ is
increased up to $3\pi$, another `$0$-$\pi$ like'
transition is found at $\theta_B \approx 2.1\pi$

As shown in Fig.\ \ref{fig7:N4AJE}(b), the finite supercurrent
at $\varphi =0$ is induced by the interplay between
SO interaction and Zeeman effect. $I(\varphi =0)$ is plotted
as a function of $\theta_B$ in Fig.\ \ref{fig8:N4CC}(b).
The anomalous current decreases first, and
sharply increases in the `$\pi$ like'-state region.
This behavior is qualitatively the same as that for
$N=1$ in Fig.\ \ref{fig6:N1CC}(b). However,
the maximum of $|I(0)|$ is smaller.

Besides the anomalous Josephson current,
the direction dependence of critical current is
observed in the case of $N>1$.
Figure \ref{fig8:N4CC}(c) shows $I_{{\rm c},\pm}$ as
a function of $\theta_B$. Both critical currents
oscillate with the cusps at the local minima of $I_{{\rm c},\pm}$.
The position of cusps also depends on the current direction.
In Fig.\ \ref{fig8:N4CC}(c), $I_{{\rm c},+}$ and $I_{{\rm c},-}$
show the cusps below and above the critical points of
transition in Fig.\ \ref{fig8:N4CC}(a), respectively.

\begin{figure}
\includegraphics[width=60mm]{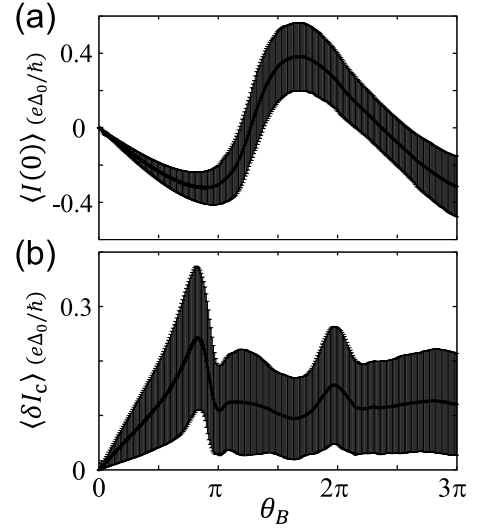}
\caption{
Calculated results of random average when $N=4$ and $l_0/L=1$.
The SO interaction is $k_\alpha /k_{\rm F} =0.15$.
The magnetic field is applied to the $y$ direction.
The random average is taken for 400 samples.
(a) Average of anomalous Josephson current,
$\langle I(\varphi =0) \rangle$ as a function of magnetic field,
$\theta_B$. Error bars represent the average of fluctuation,
$\sqrt{\langle [\Delta I(0)]^2 \rangle}$, where
$\Delta A \equiv A - \langle A \rangle$.
(b) Average of difference of critical current,
$\langle \delta I_{\rm c} \rangle$, where
$\delta I_{\rm c} \equiv |I_{{\rm c},+}-I_{{\rm c},-}|$.
Error bars are
$\sqrt{\langle [\Delta \left\{ \delta I_{\rm c} \right\}]^2 \rangle}$.
}
\label{fig9:N4AVE}
\end{figure}

In Fig.\ \ref{fig9:N4AVE}, we examine the random average
of current regarding impurity potentials. The number of
samples is 400. Figure \ref{fig9:N4AVE}(a) shows the average
of anomalous supercurrent, $\langle I(0) \rangle$,
with its fluctuation, $\sqrt{\langle [\Delta I(0)]^2 \rangle}$.
The average of anomalous current indicates negative and
positive values alternatively as a function of $\theta_B$,
where $|\langle I(0) \rangle|$ is enlarged up to about
$0.4 e\Delta_0 /\hbar$. On the other hand,
$\sqrt{\langle [\Delta I(0)]^2 \rangle}$ is saturated at
about $0.15 e\Delta_0 /\hbar$. The inversion of sign of
$\langle I(0) \rangle$ attributes to the `$0$-$\pi$ like' transition.
Roughly speaking, the current phase relation transits
from $I(\varphi) \sim \sin (\varphi - \varphi_0)$ to
$-\sin (\varphi - \varphi_0)$. Then, $I(0)$ changes
the sign from negative to positive for $\varphi_0 >0$.

In Fig.\ \ref{fig9:N4AVE}(b), we consider the average of
$\delta I_{\rm c} \equiv |I_{{\rm c},+} - I_{{\rm c},-}|$.
In the absence of magnetic field, the critical current for
positive and negative direction is equal to each other,
$\langle \delta I_{\rm c} \rangle =0$.
$\langle \delta I_{\rm c} \rangle$ increases with
increase of $\theta_B$.
$I_{{\rm c},\pm}$ sharply changes around the cusps.
Thus $\langle \delta I_{\rm c} \rangle$ becomes maximum
around the `$0$-$\pi$ like' transition. In the case of $N>1$,
the oscillation of critical current is strongly affected by
the impurity scattering. Then, the average and its fluctuation
are saturated and almost constant for $\theta_B >\pi$ except
the vicinity of critical points of transition.

\section{CONCLUSIONS AND DISCUSSION}

We have studied numerically the DC Josephson effect in
quasi-one-dimensional semiconductor nanowire with
strong SO interaction when the Zeeman effect is present.
We have examined the tight-binding model to describe
the electron and hole transport in the normal region in
S/NW/S junction. The magnetic field and Rashba SO interaction
are considered in the normal region. We have focused on
the case of short junction, where the length of
normal region is much smaller than the coherent length, $L \ll \xi$.
In the absence of SO interaction, the Andreev levels are
invariant against the inversion of phase difference $\varphi$
between two superconductors. As a result, the Josephson
current satisfies $I(-\varphi) = -I(\varphi)$, where
no supercurrent is obtained at $\varphi =0$.
The $0$-$\pi$ transition accompanying an oscillation of
critical current is observed when the magnetic field is increased.
We have introduced a parameter $\theta_B$ for magnetic field,
which describes the spin-dependent phase shift of
electron and hole transport in the normal region.
At $\theta_B \approx (2m+1)\pi/2$, the $0$-$\pi$ transition
takes place and the cusp of critical current is found.
In the presence of Rashba interaction,
we have demonstrated the anomalous Josephson effect.
The Andreev levels does not keep the relation of
$E_n (-\varphi) = E_n (\varphi)$ when the magnetic field is applied.
As a result, the phase difference $\varphi_0$ at
the minimum of ground-state energy is deviated from
$0$ and $\pi$ ($\varphi_0$-state). The current phase
relation becomes $I(-\varphi) \ne -I(\varphi)$, where
the anomalous supercurrent at $\varphi =0$ is obtained.
In addition, the critical current depends on its current
direction when more than one conduction channel is
present in the nanowire. The critical current oscillates
as a function of $\theta_B$, where the position of
cusps also depends on the current direction.
The transition between $\varphi \approx 0$ and
$\varphi \approx \pi$ takes place between
the cusps of positive and negative currents.

\begin{figure}
\includegraphics[width=85mm]{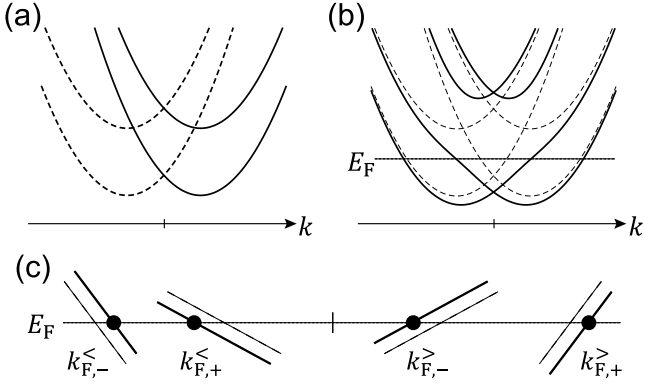}
\caption{
Schematic views of dispersion relation with SO interaction.
(a) Spin-splitting due to the $p_x \sigma_y$ term in
Eq.\ (\ref{eq:Rashba}). Solid and broken lines indicate
the branches with spin $\sigma =+$ and $-$, respectively.
(b) Dispersion relation mixed by $p_y \sigma_x$ term (solid line).
The broken line indicate the case without $p_y \sigma_x$ term.
(c) Shift of wavenumber due to the Zeeman effect in
the vicinity of Fermi level.
}
\label{fig12:Disppersion}
\end{figure}

Our calculated results have exhibited the anomalous supercurrent,
$I(\varphi =0) \ne 0$, and the direction-dependence of
critical current, $I_{{\rm c},+} \ne I_{{\rm c},-}$, when $N>1$,
in accordance with the single scatterer model.~\cite{YEN}
$I(\varphi =0) \ne 0$ is found even when $N=1$, which
points out a role of the spin-dependent Fermi velocity on
the anomalous Josephson effect.
Krive {\it et al.} also have discussed the role of Fermi velocity
in the case of long junction.~\cite{Krive1}
The dispersion relation with SO interaction is
schematically shown in Fig.\ \ref{fig12:Disppersion}.
In the nanowire, electrons are confined in
the $y$ direction. Thus, the $p_x \sigma_y$ term
in Rashba interaction in Eq.\ (\ref{eq:Rashba}) mainly
contributes to the dispersion relation rather than
the $p_y \sigma_x$ term. The $p_x \sigma_y$ term
induces the spin-splitting at $k \ne 0$ [Fig.\ \ref{fig12:Disppersion}(a)].
The spins are directed in the $\pm y$ directions.
The term of $p_y \sigma_x$ mixes the lowest branch with
spin $\sigma =\pm$ and the second lowest one with $-\sigma$.
Due to the mixing, the Fermi velocity depends on the spin direction,
as shown in Fig.\ \ref{fig12:Disppersion}(b).
Here, $\sigma =\pm$ is not good quantum number.
However, since the spins are almost directed to
the $y$ axis, we use $\sigma =\pm$ to indicate the spin.
We focus on the vicinity of Fermi energy in
Fig.\ \ref{fig12:Disppersion}(c).
When the magnetic field is applied in the $y$ direction,
the branches with spin $\sigma =\pm$ go downward and upward,
respectively. The wavenumbers are modified spin-dependently,
$k_{{\rm F},\pm}^> = k_{\rm F} \pm k_\alpha \pm E_{\rm Z}/(\hbar v_{{\rm F},\pm})$
for positive wavenumber and
$k_{{\rm F},\pm}^< =-k_{\rm F} \pm k_\alpha \mp E_{\rm Z}/(\hbar v_{{\rm F},\pm})$
for negative one.
Although $k_{\rm F} \pm k_\alpha$ is also modified by
the channel mixing, that does not affect the following discussion.
By applying these spin-dependent shifts of wavenumber,
$k_{{\rm F},\pm}^>$ and $k_{{\rm F},\pm}^<$, to $\hat{\tau}_B$ in
the single scatter model (see Appendix), the Andreev levels for
$N=1$ are given as
\begin{eqnarray}
E_{\uparrow \pm}(\varphi) &=& \Delta_0
\cos \left[
 \frac{\theta_B}{2} + \arccos \left( \pm
\sqrt{ \frac{ 1 + \delta_B + T \cos ( \varphi + \varphi_0 )}{2}}
\right) \right],
\label{eq:ALup1}
\\
E_{\downarrow \pm}(\varphi) &=& \Delta_0
\cos \left[
-\frac{\theta_B}{2} + \arccos \left( \pm
\sqrt{ \frac{ 1 + \delta_B + T \cos ( \varphi + \varphi_0 )}{2}}
\right) \right],
\label{eq:ALdown1}
\end{eqnarray}
where $\theta_B =L
(k_{{\rm F},+}^> - k_{{\rm F},+}^< - k_{{\rm F},-}^> + k_{{\rm F},-}^<) /2$,
$\delta_B = (1-T) \cos \left\{ \theta_B (2x_0 -L)/L \right\}$, and
\begin{eqnarray}
\varphi_0 = \frac{L}{2}
(k_{{\rm F},+}^> + k_{{\rm F},+}^< + k_{{\rm F},-}^> + k_{{\rm F},-}^<)
= E_{\rm Z} L \left(
\frac{1}{\hbar v_{{\rm F},+}} - \frac{1}{\hbar v_{{\rm F},-}}
\right).
\label{eq:varphi0}
\end{eqnarray}
$T$ is a transmission probability of the scatterer at $x=x_0$
without the SO interaction.
This $\varphi_0$ is proportional to the magnetic field,
and results in the anomalous Josephson effect.
If the $p_y \sigma_x$ term in Eq.\ (\ref{eq:Rashba})
is disregarded, we find no anomalous Josephson effect
even when $N>1$ (not shown).
In Ref.\ \onlinecite{YEN}, the single scatterer model
demonstrated the anomalous current and
the direction-dependence of critical current when $N>1$.
Then, the single scatterer mixes the conduction channels
spin-dependently, which effectively plays the same role as
the spin-dependent Fermi velocity in the electron transport.

In this paper, we have assumed $k_\alpha /k_{\rm F} =0.15$.
Typical value of Rashba constant in experiments is
$\alpha = 3-4\times 10^{-11}\, \mathrm{eV \cdot m}$
for InAs or InGaAs.~\cite{Nitta,Grundler,Yamada}
The SO interaction in InSb tend to be stronger than
that in InAs. For $m^* =0.014 m_e$ (InSb) and
$\lambda_{\rm F}=90\, \mathrm{nm}$,
the parameter $k_\alpha /k_{\rm F} \simeq  0.15$
corresponds to
$\alpha = 5.7\times 10^{-11}\, \mathrm{eV \cdot m}$.

In recent experiments for InSb nanowire,
the direction-dependence of critical current was observed
in the magnetic field along the nanowire.~\cite{private}
This situation disagrees with our results considering
the Rashba interaction. Thus an actual SO interaction in
the nanowire is not expressed as Eq.\ (\ref{eq:Rashba}).
In the nanowire, the direction of spin quantization axis
due to the SO interaction may depend on the position $x$.
However, our discussion can be extended to the case of
general SO interaction since the anomalous Josephson effect
is observed when an applied magnetic field has a parallel
component to the spin quantization axis.
In the experiments, a few channels may exist in the nanowire.
The spacing between two superconductors is
$L \simeq 500-1000\, \mathrm{nm}$, whereas the coherent
length in the nanowire is estimated as
$\xi \sim 350\, \mathrm{nm}$. This means $L \gtrsim \xi$.
We have exhibited the anomalous Josephson effect even for
$L \ll \xi$. Therefore, the long (or intermediated-length) junction
is not essential condition. For the measurement, however,
the long nanowire is reasonable since $\varphi_0$ in
Eq.\ (\ref{eq:varphi0}) is larger as the length $L$ is longer.
The spin relaxation length due to the SO interaction is
estimated as $l_{\rm SO} \sim 200\, \mathrm{nm}$ ($\lesssim L$).
Therefore the effect of SO interaction on the Josephson effect
can be observed in experiments. The position of first cusp is
located at $\theta_B \sim \pi /2$, which corresponds to
$B \sim 0.2\, \mathrm{T}$ in our situation.
This order of magnitude is reasonable for the experiments.

\section*{ACKNOWLEDGMENT}

We acknowledge fruitful discussions about experiments
with Prof.\ L.\ P.\ Kouwenhoven, Mr.\ V.\ Mourik,
Mr.\ K.\ Zuo in Delft University of Technology, and
Assistant Prof.\ S.\ M.\ Frolov in University of Pittsburgh.

\appendix

\section{Calculation Method of Tight-Binding Model}

Here, we explain a calculation method of scattering matrix
using the Green's function.~\cite{Ando}

We consider the matrix representation of Hamiltonian in
Eq.\ (\ref{eq:tbHamiltonian}),
\begin{equation}
H = \left( \begin{array}{ccccc}
\tilde{H}_0      &-t\tilde{T}_{0,1}&                   &                            &                            \\
-t\tilde{T}_{1,0}&   \tilde{H}_1   &-t\tilde{T}_{1,2}&                           &                            \\
                   &-t\tilde{T}_{2,1}&     \ddots     &                            &                            \\
                   &                   &                    &     \tilde{H}_{N_x}      &-t\tilde{T}_{N_x,N_x+1}\\
                   &                   &                    &-t\tilde{T}_{N_x+1,N_x}&   \tilde{H}_{N_x+1}     \\
\end{array} \right),
\label{appen:H}
\end{equation}
where $\tilde{H}_j$ is a $2N_y \times 2N_y$ matrix describing
the $j$-th slice,
\begin{equation}
\tilde{H}_j =
\left( \begin{array}{cccc}
(v_{j,1}+4t)\hat{1}&-t\hat{T}_{j,1;j,2} &         &                      \\
-t\hat{T}_{j,2;j,1} &(v_{j,2}+4t)\hat{1}&         &                      \\
                      &                      &\ddots &                      \\
                      &                      &         &(v_{j,N_y}+4t)\hat{1} \\
\end{array} \right).
\end{equation}
$\hat{1}$ is a $2\times 2$ unit matrix and $v_{j,l}$ denotes
the on-site potential at ($j,l$). The hopping term in Eq.\ (\ref{appen:H}) is
\begin{equation}
\tilde{T}_{j,j\pm 1} =
\left( \begin{array}{ccc}
\hat{T}_{j,1;j\pm 1,1}&         &                                 \\
                          &\ddots &                                 \\
                          &          &\hat{T}_{j,N_y;j\pm 1,N_y} \\
\end{array} \right).
\end{equation}

In an ideal lead, the wavefunctions of conduction channels
are written as
\begin{eqnarray}
\psi_\mu (j,l) &=& \exp( ik_{\mu}a j) u_{\mu} (l), \\
   u_{\mu} (l) &=& \sqrt{\frac{2a}{W}} \sin \left( \frac{\pi \mu la}{W} \right).
\label{appn:Wavefunc}
\end{eqnarray}
The wavenumber $k_{\mu}$ satisfies
$E_\mu (k_\mu ) = E_{\rm F}$, where
the dispersion relation is given by
\begin{equation}
E_\mu (k)= 4t - 2t\cos
\left( \frac{\pi \mu a}{W} \right) -2t\cos (k a).
\end{equation}
The band edge, $E_\mu (k=0)$, is located below
$E_{\rm F}$ for the conduction modes.
The wavefunction of evanescent mode is written as
\begin{equation}
\psi_\mu (j,l) = \exp (-\kappa_{\mu}a j) u_{\mu} (l).
\end{equation}
The band edge is located above $E_{\rm F}$ and $\kappa_{\mu}$
is determined from $E_\mu (i \kappa_{\mu})=E_{\rm F}$.
Here, we introduce some matrices for the calculation of
scattering matrix. $U=(\bm{u}_1, \bm{u}_2, \cdots, \bm{u}_{N_y} )$
is an unitary matrix, with
$\bm{u}_{\mu}=(u_{\mu}(1), u_{\mu}(2) , \cdots , u_{\mu}(N_y))^{\rm T}$
in Eq.\ (\ref{appn:Wavefunc}).
$\Lambda={\rm diag}(\lambda_1,\lambda_2,\cdots,\lambda_{N_y})$,
where $\lambda_{\mu}=\exp ({\rm i}k_{\mu} a)$ for conduction channels
and $\lambda_{\mu}=\exp (-\kappa_{\mu}a)$ for evanescent modes.

The retarded Green's function is defined as
\begin{equation}
G = \frac{1}{EI - H + \Sigma},
\end{equation}
where $\Sigma$ is the self-energy representing
the coupling with leads,
\begin{equation}
\Sigma =
\left( \begin{array}{ccc}
tF(-)^{-1} & &       \\
             & &       \\
             & &tF(+) \\
\end{array} \right)
\end{equation}
with $F(\pm ) = U \Lambda^{\pm 1} U^{-1}$.
The Green's function connects the amplitudes of
incoming and outgoing waves at the slices
$j =0, N_x +1$,
\begin{widetext}
\begin{equation}
\left( \begin{array}{c}
\bm{C}_0         (-) \\
\bm{C}_{N_x +1} (+)
\end{array} \right)
=
\left( \begin{array}{cc}
       -t G_{0,0} [F^{-1}(+) - F^{-1}(-)] -1&        -t G_{0,N_x +1} [F(+) - F(-)]    \\
-t G_{N_x +1,0} [F^{-1}(+) - F^{-1}(-)]    & -t G_{N_x +1,N_x +1} [F(+) - F(-)]-1 \\
\end{array} \right)
\left( \begin{array}{c}
\bm{C}_0         (+) \\
\bm{C}_{N_x +1} (-)
\end{array} \right),
\label{appen:matrix}
\end{equation}
\end{widetext}
where $G_{j,j^\prime}$ is the $2N_y \times 2N_y$ matrix for
the $(j,j^\prime )$ component of $G$. The vectors $\bm{C}_0 (\pm )$
and $\bm{C}_{N_x +1} (\pm )$ yield the coefficients of waves of
conduction channels in the ideal leads [Fig.\ \ref{fig1:Model}(b)] as
\begin{eqnarray}
\bm{a}_{\rm eL} &=& \sqrt{V} U^{-1} \bm{C}_0 (+), \\
\bm{a}_{\rm eR} &=& \sqrt{V} U^{-1} \bm{C}_{N_x +1} (-), \\
\bm{b}_{\rm eL} &=& \sqrt{V} U^{-1} \bm{C}_0 (-), \\
\bm{b}_{\rm eR} &=& \sqrt{V} U^{-1} \bm{C}_{N_x +1} (+),
\end{eqnarray}
where $\sqrt{V} \equiv {\rm diag} (\sqrt{v_{{\rm F},1}},
\cdots ,\sqrt{v_{{\rm F},N_y}})$
is a diagonal matrix of square root of velocities in
Eq.\ (\ref{eq:velocity}). If the channel is not conductive,
the velocity is zero. By substituting these equations to
Eq.\ (\ref{appen:matrix}), the scattering matrix
$\hat{S}_{\rm e}$ in Eq.\ (\ref{eq:Smatrix}) is obtained.

\section{Single Scatterer Model}

In this appendix, we explain the single scatterer
model in Ref.\ \onlinecite{YEN}.
The nanowire is along the $x$ axis and connected with
two superconductors at $x<0$ and $x>L$.
The pair potential is induced in the nanowire,
the absolute value of which is constant,
$\Delta (x) = \Delta_0 e^{i\varphi_{\rm L}}$ at $x<0$ and
$\Delta_0 e^{i\varphi_{\rm R}}$ at $L<x$, whereas
$\Delta (x) = 0$ in the normal region at $0<x<L$.
We consider the short junction, $L \ll \xi$.
A single scatterer describing an elastic scattering due to
impurities and the SO interaction in the nanowire is introduced
at $x=x_0$.

The scattering matrix for electrons by the scatterer is
denoted as $\hat{S}_{\rm scatt}$, which is given by
the matrix of orthogonal ensemble in the absence of
SO interaction and that of symplectic ensemble in
the limit of strong SO interaction.

\begin{figure}
\includegraphics[width=75mm]{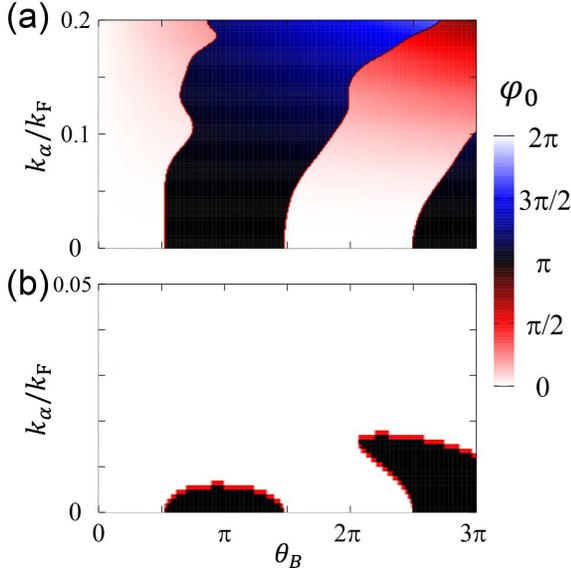}
\caption{(Color online)
Grayscale plot of the phase difference $\varphi_0$ at
the minimum of ground-state energy in the plane of
magnetic field $\theta_B$ and SO interaction
$k_\alpha /k_{\rm F}$ when $N=1$ and $l_0/L=1$.
The sample is same as that in Fig.\ \ref{fig5:N1AJE}.
The magnetic field is applied in the $y$ (a) and $x$ directions (b).
}
\label{fig10:cPhase}
\end{figure}

In the presence of magnetic field, the Zeeman effect
is taken into account as the spin-dependent phase shift
for electrons and holes in the propagation through
the normal region. The Zeeman effect shifts
the wavenumber as
$k_\pm^>= k_{\rm F} +(E \pm E_{\rm Z})/(\hbar v_{\rm F})$
for $k>0$ and
$k_\pm^<=-k_{\rm F} -(E \pm E_{\rm Z})/(\hbar v_{\rm F})$
for $k<0$.~\cite{Chtchelkatchev1}
The propagation of electron with spin $\sigma =\pm$ and
hole with $\sigma =\mp$ at $0<x<x_0$ acquires the phase
$\pm \theta_{B{\rm L}} =\pm 2E_{\rm Z}x_0/(\hbar v_{\rm F})$,
whereas that at $x_0<x<L$ is
$\pm \theta_{B{\rm R}} =\pm 2E_{\rm Z}(L-x_0)/(\hbar v_{\rm F})$.
Here, $v_{\rm F}$ is independent of channels. The terms of
$2Ex_0 /(\hbar v_{\rm F})$ and $2E(L-x_0) /(\hbar v_{\rm F})$
are safely disregarded for short junctions. The terms of
$k_{\rm F}$ are canceled out by each other.
The phases are represented by the scattering matrix,
\begin{equation}
\hat{\tau}_B = \left( \begin{array}{cc}
 \hat{1} \otimes \hat{\tau}_{B {\rm L}} & \\
 & \hat{1} \otimes \hat{\tau}_{B {\rm R}}
\end{array} \right)
\label{eq:tauB}
\end{equation}
with
\begin{equation}
\hat{\tau}_{B {\rm L(R)}} = \left( \begin{array}{cc}
e^{i \theta_{B {\rm L (R)}} /2} & \\
 & e^{-i \theta_{B {\rm L (R)}} /2}
\end{array} \right).
\end{equation}
$\hat{1}$ is an $N \times N$ unit matrix.
We fix an asymmetric parameter
$\alpha_B \equiv x_0/ (L-x_0) =\sqrt{2}$.

The Andreev reflection at $x=0,L$ is also expressed in
the term of scattering matrices $\hat{r}_{\rm he}$ and
$\hat{r}_{\rm eh}$ in Sect.\ \ref{sec:SMA}.

The product of the scattering matrices yields
\begin{equation}
\det \left(\hat{1} - \hat{\tau}_B \hat{r}_{\rm eh}
\hat{\tau}^*_B \hat{S}_{\rm scatt}^* \hat{\tau}^*_B
\hat{r}_{\rm he} \hat{\tau}_B \hat{S}_{\rm scatt} \right) =0,
\label{appn:det}
\end{equation}
which determines the Andreev levels $E_n (\varphi)$.

This simple model demonstrated the anomalous
Josephson effect and the direction-dependence of
critical current when $N>1$.

\section{Magnetic field in the $x$ direction: Disappearance of $0$-$\pi$ transition}

We discuss the case of magnetic field along
the $x$ direction, which is almost perpendicular to
the spin quantization axis due to the Rashba interaction.
In this appendix, we consider only
a single conduction channel.

\begin{figure}
\includegraphics[width=65mm]{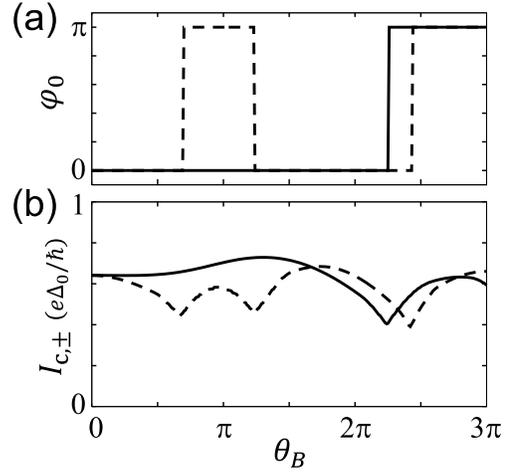}
\caption{
Calculated results for $N=1$ and $l_0/L=1$.
The magnetic field is applied to the $x$ direction.
The sample is same as that in Fig.\ \ref{fig5:N1AJE}.
(a) Phase difference $\varphi_0$ at the minimum of
ground-state energy as a function of magnetic field, $\theta_B$.
(b) Critical current $I_{{\rm c},\pm}$. The current in the positive
direction $I_{{\rm c}, +}$ is identical to that in the
negative direction $I_{{\rm c}, -}$. Solid and broken lines
in each panel indicates $k_\alpha /k_{\rm F} =0.01$
and $0.005$, respectively.
}
\label{fig11:N1Bx}
\end{figure}

Figure \ref{fig10:cPhase} shows a grayscale plots of
phase difference $\varphi_0$ at the minimum of
ground-state energy in the plane of magnetic field and
SO interaction. White and black regions mean
the $0$- and $\pi$-state, respectively.
Gray region corresponds to the $\varphi_0$-state.
In Fig.\ \ref{fig10:cPhase}(a), the magnetic field is applied
in the $y$ direction, and the anomalous Josephson effect
is obtained in the gray region. The critical points of
transition is shifted to large $\theta_B$ when
the SO interaction is stronger as mentioned in
the section \ref{sec:SO1}. We find the oscillation of
critical points as a function of $k_\alpha$,
which maybe attributes to an interference due to
the SO interaction only in the normal region.

In Fig.\ \ref{fig10:cPhase}(b), only white and black
regions are found. In the absence of SO interaction,
the $\pi$-state is realized in $\pi/2 < \theta_B < 3\pi/2$.
When $k_\alpha /k_{\rm F}$ is increased, the region of
$\pi$-state is narrower. Then, the $\pi$-state vanishes at
$l_{\rm SO}/L \lesssim 1$, where the SO length
$l_{\rm SO} \equiv \pi/(2k_\alpha)$ means a distance of
$\pi$ rotation of spins due to the SO interaction.
Figure \ref{fig11:N1Bx}(a) exhibits the phase difference
$\varphi_0$ as a function of $\theta_B$.
The $\pi$-state around $\theta_B =\pi$ disappears
with increase of $k_\alpha$.
The position of cusps of $I_{{\rm c},\pm}$ is also
gradually closer to each other, and finally
the cusps vanish [Fig.\ \ref{fig11:N1Bx}(b)].

The disappearance of $\pi$-state is interpreted by
a spin precession due to the SO interaction.
When the spin quantization axis of SO interaction
is perpendicular to the magnetic field,
the spin directed to the magnetic field is rotated by
the SO interaction. For simple consideration,
we assume that the SO interaction results in
only a spin flip in electron (hole) transport.
The Zeeman effect causes the spin-dependent phase
shift though the shift of wavenumber. If the spin flip
occurs at the middle point of normal region, the phase
shift is exactly canceled out. Then, the $0$-$\pi$
transition can be quenched by the SO interaction.
Liu {\it et al.} discussed a similar effect as $\pi$-$0$
transition by the tuning of SO interaction.~\cite{Liu2}
For in-plain magnetic field, the disappearance of
$\pi$-state coincide with the anomalous Josephson effect.
In our numerical calculation, we find large anomalous
Josephson current even when the angle between
magnetic field and SO interaction is less than
$\pi/4$ (not shown). In experiments,
the spin quantization axis may not be fixed.
Thus, the anomalous Josephson effect is observed
for arbitrary direction of magnetic field.


\begin{thebibliography}{99}
\bibitem{comphi0}
The $\varphi_0$-state is used for the $0-\pi$ junctions,
where the normal and ferromagnetic metals are arranged
in parallel without SO interaction.
The ground-state energy indicates two minima at
$\varphi = \pm \varphi_0$. See Ref.\ \onlinecite{Sickinger}
and cited references therein.
\bibitem{com1}
The Hamiltonian for the holes would be given by
$-\mathcal{T} H \mathcal{T}^{-1}$, using the time
reversal operator $\mathcal{T} = -i \hat{\sigma}_y K=\hat{g} K$,
for $(\phi_{{\rm h}-},\phi_{{\rm h}+})^{\rm T}$.
In our definition of $\bm{\psi}_{\rm h}$, $\hat{g}$ appears
in the off-diagonal part in Eq.\ (\ref{eq:BdG}).
\bibitem{comSO}
Here, $E_{\rm F}$ is the Fermi energy in the absence of
SO interaction. In the presence of SO interaction,
we shift the energy by $-\hbar^2 k_\alpha^2 /(2m^*)$
to keep the electron density constant.~\cite{EtoQPC}
\bibitem{com2}
If the length of junction or the energy-dependence of
scattering matrix are taken into account, the Andreev
levels are split by the SO interaction at
$\varphi \ne 0, \pm \pi$ since a finite phase difference
breaks the time-reversal symmetry, mentioned in
Ref.\ \onlinecite{Beri}.
\end{thebibliography}
\end{document}